\documentclass[journal,10pt]{IEEEtran}
\usepackage[T1]{fontenc}
\usepackage{dsfont}
\usepackage{array}
\usepackage{booktabs}
\usepackage{amsmath}
\usepackage{bbm} 
\ifCLASSINFOpdf
\else
\fi
\usepackage{color}
\usepackage{cancel}
\usepackage{cite}
\usepackage{algorithmic}
\usepackage[ruled]{algorithm2e}
\usepackage{ulem}
\usepackage{multirow}
\usepackage{subfigure}
\usepackage{textcomp}
\usepackage{flushend}
\usepackage{amsmath}
\usepackage{amssymb}
\usepackage[cmintegrals]{newtxmath}

\usepackage{stfloats}
\usepackage{float}
\usepackage{graphicx}
\usepackage{xcolor}
\usepackage{graphics} 
\usepackage{epsfig}
\usepackage{epstopdf}

\newcommand{\RNum}[1]{\uppercase\expandafter{\romannumeral #1\relax}}

\usepackage{balance}

\begin{document}
	\title{Simultaneous Indoor and Outdoor Coverage with Conformal Intelligent Omni-Surface\\
	}  
	\author{Yiqi~Chen,~Xiaoming~She,~Jianchi~Zhu,~Nanxi~Li,~Ruizhe~Long,~Ying-Chang~Liang
    \thanks{This work has been submitted to the IEEE for possible publication. Copyright may be transferred without notice, after which this version may no longer be accessible.}


    }    

	
	\maketitle
\begin{abstract}
    Seamless indoor--outdoor coverage conventionally relies on coordinated outdoor macro cells and indoor small cells, incurring high deployment and backhaul costs. Intelligent omni-surfaces (IOSs) provide a promising alternative by enabling simultaneous reflection and transmission across building boundaries. Enabled by recent advances in conformal antennas, conformal IOSs (CIOSs) mounted on building facades offer a flexible and deployment-friendly solution for integrated indoor--outdoor coverage. This paper develops a stochastic-geometry framework to evaluate the downlink performance of CIOS-assisted networks serving both indoor and outdoor users. Specifically, a building-dependent spatial model is established where randomly located buildings are modeled via a Boolean scheme with CIOSs mounted on their facades. Unlike planar IOS deployment, conformal integration introduces geometry dependent visibility constraints, which are captured through the effective numbers of participating elements defined by visible arc metrics for reflection and transmission. Incorporating these constraints, coverage probability expressions are derived under three CIOS operating protocols. Numerical results show that, compared with planar IOS, CIOS improves the successful connection probabilities of the reflected and transmitted links by up to 75\% and 125\%, respectively, while the overall coverage gain remains dependent on blockage conditions because curvature reduces the effective serving aperture.
\end{abstract}


\vspace{-0.1cm}
\section{Introduction}
\IEEEPARstart{T}he evolution toward 6G networks demands ubiquitous, high-capacity wireless coverage that seamlessly serves both indoor users (IUs) and outdoor users (OUs) \cite{1}. However, achieving seamless indoor--outdoor coverage remains fundamentally challenging, as signals transmitted by outdoor macro base stations (BSs) experience significant penetration loss when propagating through building materials such as brick and concrete \cite{2}. Conventionally, indoor coverage gaps are addressed by densely deploying small cells. While effective for capacity, this approach incurs high capital and operational costs, primarily due to the requisite dense backhaul infrastructure \cite{3,4}. Deploying relays or repeaters near windows provides an alternative solution to extending outdoor signals into indoor environments \cite{5,6}, yet these active devices introduce complexities in installation, powering, and maintenance, limiting their practical scalability. Therefore, a cost-effective, scalable, and architecturally integrated solution is critically needed to bridge the indoor--outdoor coverage gap.

The intelligent omni-surface (IOS), also referred to as the simultaneous transmitting and reflecting reconfigurable intelligent surface (STAR-RIS), has recently emerged as a promising solution to this challenge \cite{7}. By jointly controlling the reflected and transmitted signals, an IOS can shape the wireless propagation environment for users located on both sides of the surface \cite{8}. Together with its low-profile structure, this property makes IOS particularly suitable for integration into building facades \cite{9}. In this way, building boundaries can contribute to wireless service provisioning instead of acting as propagation blockages \cite{10,11}. However, most existing studies model IOSs as two-dimensional (2D) planar structures, while the geometric coupling between IOS deployment and curved building facades remains unexplored.

Therefore, this paper introduces the concept of a conformal intelligent omni-surface (CIOS) for simultaneous indoor--outdoor services. Inspired by conformal antennas \cite{12}, a CIOS can elastically adapt to host building facades, enabling seamless integration with practical urban structures. This work develops a stochastic-geometry framework to characterize the large-scale downlink performance of CIOS-assisted urban networks and to reveal how conformal facade integration affects simultaneous outdoor reflection and indoor transmission.

\begin{figure}[!t]
	\centering	\includegraphics[width=0.4\textwidth]{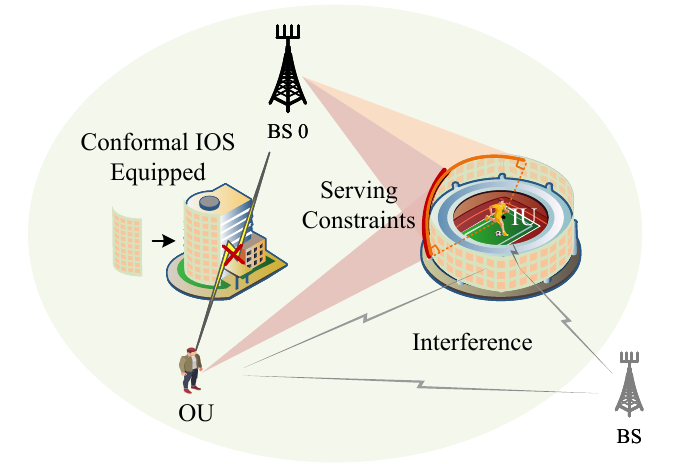}	
	\vspace{-0.4cm}
	\caption{Illustration of the system model where conformal IOSs are deployed on building facades to provide simultaneous indoor and outdoor coverage.}
	\vspace{-0.3cm}
\end{figure}

\vspace{-0.3cm}
\subsection{Related Works}

The IOS has attracted significant attention for its ability to simultaneously serve users on both sides of the surface. Existing studies have shown that IOSs can enhance signal strength, extend service range, and improve beamforming performance for both indoor and outdoor users through joint reflection and transmission design \cite{9,X2,X3}. However, these works mainly focus on link level optimization with one or a few IOSs deployed at predetermined locations, and thus do not reveal the large-scale behavior of IOS-assisted multi-cell networks.

To characterize network-level performance, stochastic geometry has become a standard analytical tool for RIS-assisted systems. Early works derived tractable coverage expressions for RIS-assisted networks \cite{X4}, and subsequent studies developed more general system-level frameworks \cite{X5,X6}. This methodology has also been extended to IOS-assisted networks, for example under non-orthogonal multiple access (NOMA) \cite{X7} or line-segment blockage models \cite{X8}. Nevertheless, these studies typically model IOSs as independently deployed points following a homogeneous Poisson point process (HPPP). Although analytically convenient, such an abstraction does not capture the fact that practical IOS deployment is often constrained by existing urban structures, especially building facades.

Motivated by this limitation, a few recent works have started to incorporate the coupling between IOS deployment and urban infrastructure. Reference \cite{X18} studies IOS-assisted connectivity over randomly distributed rectangular blockages using percolation theory. More closely related to this paper, \cite{X9} develops a stochastic-geometry framework for indoor--outdoor networks in which buildings are modeled by a Boolean scheme with cylindrical blockages, users are partitioned into indoor and outdoor groups, and each IOS is deployed as a planar surface at a specific position on the building facade. However, both \cite{X18} and \cite{X9} still adopt non-conformal planar deployment on building facades, and therefore cannot capture the effect of conformal integration with building geometry.

Another closely related research direction concerns conformal deployment. Enabled by conformal-antenna technology \cite{12} and flexible metasurface implementations \cite{X12}, curved and bendable surfaces have become increasingly relevant for practical wireless integration. From a theoretical perspective, prior works on conformal reflective RIS first examined curved deployments on specific surfaces, such as arc-shaped structures and car doors \cite{X13,X14}, with a primary focus on the reflection constraints and beam-pattern distortion induced by surface curvature. Subsequent studies extended the conformal setting to cylindrical and other curved structures \cite{X15,X16}, gradually shifting the focus from link level beam control to geometry-aware performance characterization. In this direction, \cite{X17} analytically reveals that surface curvature changes the effective illuminated area, thereby affecting communication performance. Building on this insight, \cite{X16} further incorporates the curvature dependent illuminated area into a system-level metric that captures its impact on both desired and interfering signals, and derives the corresponding coverage probability. However, extending these conformal reflective RIS principles to CIOS-assisted indoor--outdoor networks remains non-trivial, and several key gaps still persist:
\begin{itemize}
	\item The spatial distribution of IOSs is modeled as an independent HPPP in \cite{X7}, \cite{X8}, overlooking their deployment dependency on existing urban structures. Although \cite{X9} and \cite{X18} incorporate urban infrastructure into the deployment model, they still consider IOSs as planar surfaces deployed at specific facade locations rather than conformal deployments adapted to building geometry. As a result, they cannot capture the coupling between conformal facade integration and building geometry in large-scale indoor--outdoor networks.	
	\item The impact of conformal geometry on the effective illuminated area and its ensuing effect on network performance is investigated in \cite{X17} and further characterized in \cite{X16}. However, extending the conformal paradigm from reflective RIS to IOS is non-trivial, since IOS inherently partitions users into OUs and IUs with fundamentally different geometric feasibility conditions, which have not been incorporated into existing system-level analysis.
	\item For IOS-assisted indoor--outdoor communications, \cite{X18} characterizes network connectivity and \cite{X9} derives coverage expressions under non-conformal deployment. However, conformal deployment reshapes the feasible line-of-sight (LoS) path distributions of both desired and interfering cascaded links, so the existing analytical frameworks in \cite{X18} and \cite{X9} cannot be directly applied.
\end{itemize}

\vspace{-0.5cm}
\subsection{Contributions}
\vspace{-0.1cm}
To address the research gap, this paper introduces the CIOS as a new metasurface paradigm that is conformally mounted on curved building facades to simultaneously provide outdoor reflection and indoor transmission services. The resulting geometry-dependent reflection and transmission constraints arising from the curved surface are theoretically characterized, and their impact on large-scale network performance is analyzed via stochastic geometry. The main contributions are summarized as follows:
\begin{itemize}	
	\item A building dependent system model is developed to capture the spatial coupling among buildings, facade-mounted CIOSs, and outdoor BSs. Specifically, buildings are modeled by a Boolean scheme of random cylinders, each equipped with an $N$-element CIOS conformally deployed along the cylindrical facade, while outdoor BSs are modeled as a Poisson hole process restricted to the outdoor region.
    \item The geometry-induced feasible CIOS aperture for simultaneous indoor--outdoor service is explicitly characterized in a tractable form. Owing to the cylindrical conformal deployment, only a location-dependent subset of CIOS elements can participate in each cascaded link, and the feasibility conditions differ for outdoor reflection and indoor transmission. The effective number of participating elements is introduced and quantified via visible arc geometry.
	\item An analytical framework for CIOS-assisted networks is established by incorporating the conformal service constraints into the statistics of both desired and interfering cascaded links. The successful connection probabilities of the reflected and transmitted links are derived, which further lead to the coverage probabilities of OUs, IUs, and the overall network under different CIOS operating protocols. Numerical results show that: (i) compared with the planar IOS baseline, CIOS improves the successful connection probabilities of the reflected and transmitted links by up to 75\% and 125\%, respectively, by enlarging the feasible service region; (ii) the coverage gain of CIOS is blockage-dependent, since conformal deployment enhances geometric service accessibility but simultaneously reduces the effective serving aperture, leading to a tradeoff between link feasibility and cascaded-link strength.
\end{itemize}

\vspace{-0.1cm}
\section{System Model}
This paper considers a downlink indoor--outdoor communication system assisted by CIOSs. As illustrated in Fig.~1, CIOSs are conformally mounted on building facades to mitigate the high penetration loss and wall-induced blockages typical of building boundaries, thereby establishing controllable propagation paths between indoor and outdoor environments. The remainder of this section details the network model, association scheme, CIOS serving constraints, CIOS operating protocols, and the channel model. 

\vspace{-0.2cm}
\subsection{Network Model}\label{AA}
\vspace{-0.1cm}
We model randomly located buildings by a Boolean scheme of identical cylinders with fixed radius $D$ $(D>0)$. The building centers are distributed according to a HPPP $\Phi_{\rm BL}$ with intensity $\lambda_{\rm BL}$ on $\mathbb{R}^2$. The horizontal footprint of a building centered at $\mathbf{X}_i\in\Phi_{\rm BL}$ is represented by the disk $\mathcal D(\mathbf{X}_i,D)\triangleq \left\{\mathbf{x}\in\mathbb{R}^2:\|\mathbf{x}-\mathbf{X}_i\|\le D\right\}$. Accordingly, the overall building region is given by ${{\cal A}_{{\rm{BL}}}} \buildrel \Delta \over = \bigcup\nolimits_{{{\bf{X}}_i} \in {\Phi _{{\rm{BL}}}}} {{\cal D}({{\bf{X}}_i},D)}$.

In this work, the outdoor BSs provide wireless services assisted by the facade-mounted CIOS. Their locations are modeled by restricting an independent HPPP $\Phi_{\rm{BS}}$ of intensity $\lambda_{\rm{BS}}$ to the outdoor region $\mathbb{R}^2\setminus\mathcal A_{\rm BL}$, forming a Poisson hole process defined as $\Phi_{\rm{BS}}^{(0)} \triangleq \Phi_{\rm{BS}} \cap \left(\mathbb{R}^2\setminus\mathcal A_{\rm BL}\right)$. The effective density of outdoor BSs is $\lambda _{\rm{BS}}^{(0)}={\lambda _{\rm{BS}}}{e^{ - {\lambda _{\rm{BL}}}\pi {D^2}}}$, which is equivalent to thinning the original PPP $\Phi_{\rm{BS}}$ with the void probability of buildings. 

The user population is divided into two categories, IUs and OUs. IUs are distributed uniformly within each building disk. Specifically, conditioned on a building centered at $\mathbf{X}_i$, the IU locations form a uniform point process over $\mathcal D(\mathbf{X}_i,D)$ with intensity ${\lambda _{{\rm{IU}}}}$. Consequently, the distance $r_{{L_0}I} = |\mathbf{u}_i - \mathbf{X}_i|$ from an IU to its building center follows the probability distribution function (PDF) $f_{r_{{L_0}I}}(r) = \frac{2r}{D^{2}}, \; 0 \le r \le D$.

OUs are confined to the outdoor region in a manner analogous to the BS deployment. Starting from an independent HPPP $\Phi_{\text{OU}}$ with intensity ${\lambda _{{\rm{OU}}}}$, the OU point process is obtained as $\Phi_{\text{OU}}^{(0)} \triangleq \Phi_{\text{OU}} \cap \left(\mathbb{R}^{2}\setminus\mathcal A_{\rm BL}\right)$.

For the subsequent system-level analysis, we adopt the perspective of a typical user randomly selected from each population, namely a typical OU and a typical IU. Due to the stationarity and ergodicity of the underlying point processes, the performance experienced by these typical users statistically represents the average network performance for all users of their respective types.

\vspace{-0.3cm}
\subsection{Association Scheme}\label{AA}
\vspace{-0.1cm}

To exploit the characteristics of CIOSs while ensuring reliable service, we adopt the following association scheme.

\textit{Outdoor Users:} For the typical OU, the distance to its nearest BS, denoted as BS 0, is $r_{{B_0}O}$, which is characterized using the equivalent-PPP distance statistics as ${f_{{r_{{B_0}O}}}}(r) = 2\pi \lambda _{{\rm{BS}}}^{(0)}r{e^{ - \pi \lambda _{{\rm{BS}}}^{(0)}{r^2}}}$. If a LoS path exists between BS 0 and the user, the OU is served via the direct link. Otherwise, when the direct link is blocked, the OU connects to the nearest building equipped with a CIOS, denoted as CIOS 0. Since a typical OU lies in the outdoor region, ${f_{{r_{{C_0}O}}}}(r) = 2\pi {\lambda _{{\rm{BL}}}}r{e^{ - \pi {\lambda _{{\rm{BL}}}}({r^2} - {D^2})}},\;r \ge D$.
Let the angle subtended at the user by BS 0 and CIOS 0 be $\varphi_O$, which is uniformly distributed over $[0, 2\pi)$. The distance between BS 0 and CIOS 0 can therefore be expressed as $r_{{B_0}{C_0}} = \sqrt {r_{{B_0}O}^2 + r_{{C_0}O}^2 - 2{r_{{B_0}O}}{r_{{C_0}O}}\cos {\varphi _O}} $.

\textit{Indoor Users:} Since BSs are deployed exclusively outdoors, an IU must be served via the transmission from a CIOS. The IU is associated with the nearest outdoor BS to the center of its host building, denoted as BS 0. The distance between the BS and the building center, $r_{{B_0}{L_0}}$, is characterized in the same manner as
${f_{{r_{{B_0}{L_0}}}}}(r) = 2\pi \lambda _{{\rm{BS}}}^{(0)}r{e^{ - \pi \lambda _{{\rm{BS}}}^{(0)}({r^2} - {D^2})}},\quad r \ge D$,
which accounts for the fact that no BS lies inside the building. Let $\varphi_I$ be the angle at the building center formed by the IU and BS 0. Under the uniform indoor distribution, $\varphi_I \sim U[0, 2\pi)$. The resultant straight-line distance between BS 0 and the IU is then given by $r_{{B_0}I} = \sqrt {r_{{L_0}I}^2 + r_{{B_0}{L_0}}^2 - 2{r_{{L_0}I}}{r_{{B_0}{L_0}}}\cos {\varphi _I}}$.

\begin{figure}[!t]
	\centering	\includegraphics[width=0.5\textwidth]{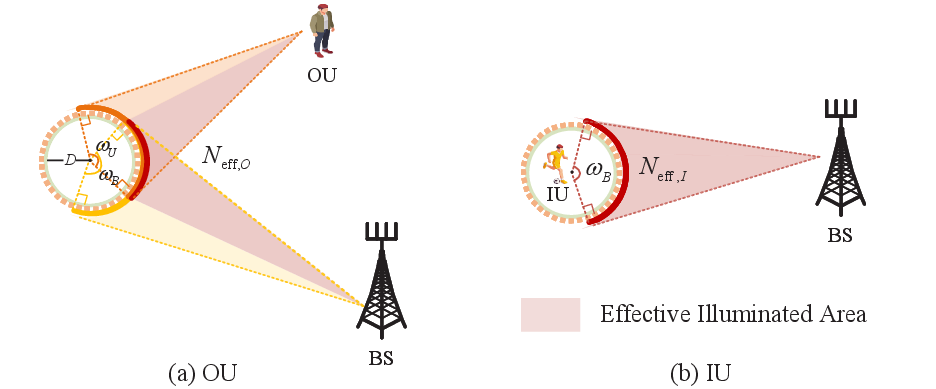}	
	\vspace{-0.8cm}
	\caption{The effective serving elements of CIOS (a) outdoor (b) indoor.}
	\label{f1}
	\vspace{-0.5cm}
\end{figure}

\vspace{-0.2cm}
\subsection{CIOS Operating Protocols}
\vspace{-0.1cm}

To support simultaneous indoor transmission and outdoor reflection, the CIOS can operate under three common IOS control protocols, namely energy splitting (ES), time switching (TS), and mode switching (MS). To unify the subsequent analysis, we introduce protocol-dependent indoor and outdoor allocation factors, denoted by $\eta_I^{p}$ and $\eta_O^{p}$ for $p\in\{{\rm ES,TS,MS}\}$. When protocol $p$ is active, $\eta_I^{p}$ and $\eta_O^{p}$ characterize the corresponding resource partition and satisfy $\eta_I^{p}+\eta_O^{p}=1$, where the allocated resource refers to power under ES, time under TS, and the portion of CIOS elements under MS.

For the two inactive protocols, their allocation factors are set to unity, i.e., $\eta_I^{q}=\eta_O^{q}=1$ for $q\neq p$, such that they do not affect the unified expressions derived later. For example, under ES, only $(\eta_I^{\rm ES},\eta_O^{\rm ES})$ is used for resource partition, while $(\eta_I^{\rm TS},\eta_O^{\rm TS})$ and $(\eta_I^{\rm MS},\eta_O^{\rm MS})$ are both set to $(1,1)$. The same convention applies to TS and MS.

Accordingly, the coverage analyses can be written in a unified form. In particular, ES affects the reflected and transmitted cascaded link powers through the corresponding power splitting factors, MS scales the effective numbers of available CIOS elements through $\eta_O^{\rm MS}$ and $\eta_I^{\rm MS}$, whereas TS does not change the geometry-driven effective aperture of an instantaneous link and enters the overall coverage only through the time allocation weights.

\vspace{-0.3cm}
\subsection{Effective Participating Elements}
\vspace{-0.1cm}
Assume each facade-mounted CIOS consists of $N$ elements uniformly deployed along the curved building perimeter. Due to the conformal geometry, only a subset of elements can effectively participate in forming a CIOS-assisted link, since the incident and outgoing wavefronts are confined by the tangency-limited visibility on the curved facade. Unlike a planar surface where all $N$ elements may contribute, the effective aperture depends on the relative geometry among the user, the associated BS, and the building. To capture this effect in a tractable manner, we define the effective number of participating elements, denoted by $N_{\text{eff},O}$ and $N_{\text{eff},I}$.

\textit{OUs Reflection Link:} For an OU served via a reflection path, only the elements simultaneously visible to both the OU and its associated BS can contribute. Let $\mathcal{V}_U\subset[0,2\pi)$ and $\mathcal{V}_B\subset[0,2\pi)$ denote the angular-visible arc intervals on the building perimeter observed from the OU and the BS, respectively, with corresponding measures $\omega_U(r_{{C_0}O})\triangleq|\mathcal{V}_U|$ and $\omega_B(r_{{B_0}{C_0}})\triangleq|\mathcal{V}_B|$. The effective reflection aperture is determined by the overlap of these intervals.

\emph{Definition (Effective elements for an OU $N_{\text{eff},O}$)}:
For an OU assisted by a reflective CIOS link, the effective number of elements is
\begin{equation}
N_{\text{eff},O}
= \frac{|\mathcal{V}_U\cap\mathcal{V}_B|}{2\pi}N.
\end{equation}
If $N_{\text{eff},O}=0$, the reflection link is geometrically infeasible even when blockage is absent.

\textit{IUs Transmission Link:} For an IU served via a transmission path, the dominant geometric constraint is imposed by the BS illumination on the exterior facade. Specifically, only the facade arc bounded by the two tangents drawn from the associated BS to the building can be effectively excited and contribute to transmission.

\emph{Definition (Effective elements for an IU $N_{\text{eff},I}$)}:
For an IU served by a CIOS transmission link, the effective number of elements is determined by the illuminated arc as
\begin{equation}
N_{\text{eff},I}
= \frac{\omega_B(r_{{B_0}{L_0}})}{2\pi}N,
\end{equation}
where $\omega_B(r_{{B_0}{L_0}})$ denotes the central angle of the facade arc defined by the two tangents drawn from the BS to the building.

\emph{Remark:} $N_{\text{eff},O}$ and $N_{\text{eff},I}$ characterize the geometry-induced visibility constraint of the CIOS aperture. For interference characterization, an analogous quantity (e.g., $N_{\text{eff},O}'$) can be defined according to the corresponding interferer--receiver geometry. Among the considered protocols, MS further partitions the aperture into transmission and reflection sub-arrays, so the effective number of available elements is scaled by $\eta_I^{\rm MS}$ or $\eta_O^{\rm MS}$, whereas ES and TS do not alter the geometry-driven definitions of $N_{\text{eff},O}$ and $N_{\text{eff},I}$.

\vspace{-0.5cm}
\subsection{Channel Model}
\vspace{-0.1cm}
Both BSs and users are equipped with a single antenna, as commonly assumed in related works \cite{X7,X9,X16}. All BSs transmit with equal power $P_t$, which is normalized to unity for tractability. The channel consists of large-scale path loss and small-scale fading. We adopt the standard power-law path loss $\ell(r)=r^{-\alpha}$ with exponent $\alpha>2$, and assume independent Rayleigh fading for all links.

\textit{Outdoor links:} For a typical OU, denote by $r_{{B_i}O}$, $r_{{B_i}{C_k}}$, and $r_{{C_k}O}$ the BS--OU, BS--CIOS, and CIOS--OU distances, respectively. The corresponding baseband channels satisfy ${h_{{B_i}O}} = \sqrt {{g_{{B_i}O}}\ell ({r_{{B_i}O}})} {\mkern 1mu}$, $h_{{B_i}{C_k}}^{(n)} = \sqrt {g_{{B_i}{C_k}}^{(n)}\ell ({r_{{B_i}{C_k}}})} {\mkern 1mu}$, and $h_{{C_k}O}^{(n)} = \sqrt {g_{{C_k}O}^{(n)}\ell ({r_{{C_k}O}})} {\mkern 1mu}$, where $\{ {g_{{B_i}O}},g_{{B_i}{C_k}}^{(n)},g_{{C_k}O}^{(n)}\} \sim\exp (1)$. Let ${\Theta}_O^{k}=\mathrm{diag}\!\left(e^{j\theta_O^{k,1}},\ldots,e^{j\theta_O^{k,N}}\right)$ denote the reflection phase-shift matrix of CIOS $k$. The cascaded channel from BS $i$ to the OU via CIOS $k$ is
\begin{equation}
h_{{B_i}{C_k}O}=\big[\mathbf{h}_{{B_i}{C_k}}^{(o)}\big]^T {\Theta}_O^{k}\mathbf{h}_{{C_k}O}
=\sum_{n=1}^{N} h_{{B_i}{C_k}}^{(n)}h_{{C_k}O}^{(n)}e^{j\theta_O^{k,n}} .
\end{equation}
With ideal passive beamforming at the serving CIOS, the phases are aligned over the effective aperture, yielding
		\begin{equation}
			|h_{{B_0}{C_0}O}^{}| = \sum_{n=1}^{N_{\text{eff},O}} | h_{{B_0}{C_0}}^{\left( n \right)}| |h_{{C_0}O}^{\left( n \right)}|.
		\end{equation}

Since each product term follows a double-Rayleigh distribution, by the central limit theorem,
\begin{equation}
\begin{split}
|h_{B_0 C_0 O}| \sim \mathcal{N}\Bigg(& \frac{\pi N_{\mathrm{eff},O}}{4} \sqrt{\ell(r_{B_0 C_0}) \ell(r_{C_0 O})}, \\
&\quad \Big(1 - \frac{\pi^2}{16}\Big) N_{\mathrm{eff},O} \ell(r_{B_0 C_0}) \ell(r_{C_0 O}) \Bigg).
\label{eq:signal_gaussian}
\end{split}
\end{equation}

For interfering reflection links without phase alignment, we use $N_{\mathrm{eff},O}'$ to denote the effective number of elements and model
		\begin{equation}
			h_{{B_i}{C_k}O} \!\!=\!\!\!\! \sum_{n=1}^{N_{\text{eff},O}'}\!\!\! {h_{{B_i}{C_k}}^{(n)}} {h_{{C_k}O}^{(n)}} \!\!\mathop \sim\limits^{approx} \!\!{\cal C}{\cal N}\!\left(\! {0,{N_{\text{eff},O}'}{\ell\! \left( {{r_{{B_i}{C_k}}}} \right)}{\ell\! \left( {{r_{{C_k}O}}} \right)}} \!\right).
		\end{equation}

\textit{Indoor links:} A precise transmission model involves the two-hop attenuation over the BS--CIOS and CIOS--IU segments and requires averaging over the random IU location inside the building. To maintain tractability, we approximate the large-scale attenuation of the serving transmission link by an equivalent distance \cite{X20}
\begin{equation}
\tilde r_{{B_0}I}\triangleq r_{{B_0}{L_0}}+r_{{L_0}I},\qquad \ell(\tilde r_{{B_0}I})=\tilde r_{{B_0}I}^{-\alpha},
\end{equation}
which captures the dominant distance-dependent attenuation while avoiding an additional inner integral over the indoor location. With ideal passive beamforming, the effective channel magnitude is approximated as
\begin{equation}
	|{h_{{B_0}{C_0}I}}|\sim\mathcal{N}\left( \frac{N_{\text{eff},I} \sqrt{\pi {\ell({\tilde r_{{B_0}I}})}}}{2}, \frac{(4 - \pi) N_{\text{eff},I} {\ell({\tilde r_{{B_0}I}})}}{4} \right).
\end{equation}

Similarly, for interfering transmission links we denote the effective aperture by $N_{\mathrm{eff},I}'$ and model the equivalent complex gain as
		\begin{equation}
			h_{{B_i}{C_k}I} \mathop \sim\limits^{approx} {\cal C}{\cal N}\left( {0,{N_{\text{eff},I}'}{\ell({\tilde r_{{B_i}I}})}} \right).
		\end{equation}
where $\tilde r_{{B_i}I}$ is defined analogously for the interferer--IU geometry.

\vspace{-0.1cm}
\section{Effective Number of Serving Elements}
\vspace{-0.1cm}
Although a facade-mounted CIOS comprises $N$ reconfigurable elements, only a geometry-dependent subset can effectively contribute to a CIOS-assisted link due to the curvature-induced visibility limitation, as illustrated in Fig.~2. To quantify this effect, we focus on the effective number of participating elements, denoted by $N_{\text{eff},\xi}$ for $\xi\in\{O,I\}$. Conditioned on the serving geometry, we next characterize $N_{\text{eff},O}$ for reflection to OUs and $N_{\text{eff},I}$ for transmission to IUs, which will be used in the subsequent coverage analysis.

\emph{Lemma 1}. For an OU served via a CIOS reflection link, the PDF of $N_{\text{eff},O}$ is given by
	\begin{equation}
		\begin{array}{l}
			f_{N_{{\rm{eff}},O}}(n)
			= \dfrac{\left| \omega_U-\omega_B \right|}{2\pi}\,
			\delta\!\left(n-\dfrac{\min(\omega_U,\omega_B)\eta_O^{\rm MS}N}{2\pi}\right)\\[0.5ex]
			\quad +\dfrac{2}{\eta_O^{\rm MS}N}\,
			{\bf 1}_{\left(0,\ \frac{\min(\omega_U,\omega_B)\eta_O^{\rm MS}N}{2\pi}\right)}(n)
			+\left(1-\dfrac{\omega_U+\omega_B}{2\pi}\right)\delta(n),
		\end{array}
	\end{equation}
where $\omega_B = 2\arccos\!\left(\frac{D}{r_{{B_0}{C_0}}}\right)$, $\omega_U = 2\arccos\!\left(\frac{D}{r_{{C_0}O}}\right)$ denote the visible central angles of the building arc observed from the serving BS and the OU, respectively. Here, $\delta(\cdot)$ denotes the Dirac delta function, and ${\bf 1}_{(a,b)}(x)$ denotes the indicator function, which equals one for $x\in(a,b)$ and zero otherwise.

\begin{IEEEproof}
Please refer to Appendix A.
\end{IEEEproof}

\emph{Lemma 2}. For an IU served via a CIOS transmission link, $N_{\text{eff},I}$ is given by
\begin{equation}
N_{\text{eff},I} = \frac{\eta_I^{\rm MS} N}{\pi} \arccos\!\left( \frac{D}{r_{{B_0}{L_0}}} \right).
\label{eq:NeffI}
\end{equation}

\begin{IEEEproof}
For transmission, the effective aperture is determined by the facade arc illuminated by the serving BS, whose central angle is $2\arccos(D/r_{{B_0}{L_0}})$, as shown in Fig.~2(b). Under MS, the transmission sub-array occupies a fraction $\eta_I^{\rm MS}$ of the total elements, which directly leads to \eqref{eq:NeffI}.
\end{IEEEproof}

\begin{figure}[!t]
	\centering	\includegraphics[width=0.35\textwidth]{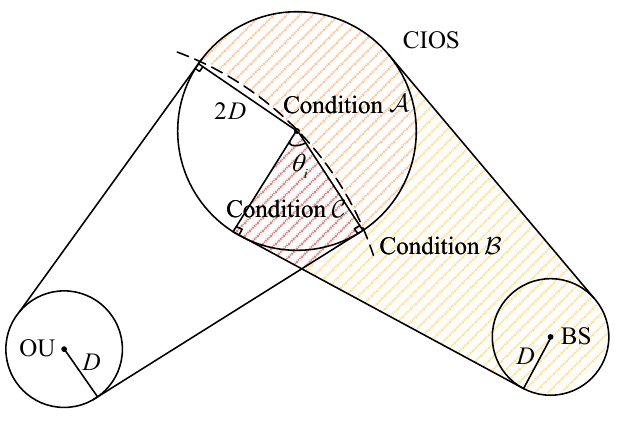}	
	\vspace{-0.3cm}
	\caption{Geometric illustration of conditions ${\cal A}$--${\cal C}$ for the BS 0-CIOS 0-OU cascaded link. The shaded regions ${\cal A}$ and ${\cal B}$ indicate the potential blockage regions for the CIOS 0--OU and BS 0-CIOS 0 links, respectively, while region ${\cal C}$ denotes the angular-overlap area where CIOS 0 is simultaneously visible to both BS 0 and the OU.}
	\label{f1}
	\vspace{-0.2cm}
\end{figure}

\vspace{-0.1cm}
\section{Probability of Successful Connection}
Due to blockage effects from buildings and the geometric constraints of the CIOS, a communication link may be blocked, whether it is the direct link between a BS and an OU, the reflective link via a CIOS, or the transmission link to an IU. To quantify the likelihood of establishing a viable connection under the considered system model, we derive the following three key probabilities:
\begin{itemize}
	\item \textit{Direct link success probability ${P_d}$:} the probability that an OU is successfully served by its nearest BS via a direct LoS link;
	\item \textit{Reflection link success probability ${P_f}$:} the probability that an OU is successfully served by a reflective link through its nearest CIOS;
	\item \textit{Transmission link success probability ${P_s}$:} the probability that an IU is successfully served by a transmission link via the CIOS on its host building.
\end{itemize}

\vspace{-0.3cm}
\subsection{Direct Link}
The OU associates with the nearest BS 0 when there are no blockages between the OU and the BS 0.

\emph{Lemma 3 (LoS Probability for Serving Direct Links)}. 
Conditioned on the serving distance $r_{{B_0}O}$, the probability that the direct BS 0--OU link is not blocked is
	\begin{equation}
		{P_d}({r_{{B_0}O}}) = \exp ( - \lambda _{\rm{BL}}(\pi {D^2} + 2D{r_{{B_0}O}})).
	\end{equation}
\begin{IEEEproof}
A building blocks the segment if its center falls in the Minkowski sum of the segment and a disk of radius $D$, whose area equals $A_1(r)=\pi D^2+2Dr$. Since building centers form a HPPP, the void probability yields $P_d(r)=\exp(-\lambda_{\rm BL}A_1(r))$. Substituting the expression for $A_1(r_{{B_0}O})$ yields the result.
\end{IEEEproof}

\emph{Corollary 1 (LoS Probability for Interfering Direct Links)}. For any interfering BS $i$ $(i \ne 0)$, the probability $P_d^I(r)$ that its direct link to the OU remains unblocked is governed by the identical geometric condition and blocking region ${A_1}\left(r\right)$ as derived for the serving direct link. Therefore, conditioned on the link distance $r$,
\begin{equation}
P_d^{I}(r)=\exp\!\big(-\lambda_{\rm BL}A_1(r)\big)=\exp\!\big(-\lambda_{\rm BL}(\pi D^2+2Dr)\big).
\end{equation}

\vspace{-0.3cm}
\subsection{Reflection Link}
\vspace{-0.1cm}
An OU is successfully served via reflection through its nearest CIOS (CIOS 0) if the following conditions occur simultaneously:
\begin{itemize}	
	\item \textit{Condition ${\cal A}$}: the CIOS 0--OU link is not blocked;
	\item \textit{Condition ${\cal B}$}: the BS 0--CIOS 0 link is not blocked;
	\item \textit{Condition ${\cal C}$}: the visible arcs of BS 0 and the OU on the facade overlap, i.e., at least one CIOS element is simultaneously visible to both.
\end{itemize}

Accordingly, the probability that a reflection link is successfully established can be written as
\begin{equation}
P_f=\mathbb{P}({\cal A}\cap{\cal B}\cap{\cal C})
=\mathbb{P}({\cal A})\!\,\mathbb{P}({\cal C})\!\,\mathbb{P}({\cal B}\mid{\cal A},{\cal C})
\approx P_{f_A}\!\,P_{f_C}\!\,P_{f_B},
\label{eq:Pf_factor}
\end{equation}
where ${\cal A}$ and ${\cal C}$ are treated as independent, since ${\cal C}$ only depends on the geometry of the serving facade, whereas ${\cal A}$ is governed by external blockage along the CIOS--OU segment.
	
\emph{Lemma 4 (LoS Probability for the CIOS 0--OU Link)}. Conditioned on $r_{{C_0}O}$, the probability that the CIOS 0--OU link is not blocked by other buildings is approximated by
\begin{equation}
P_{f_A}(r_{{C_0}O}) \approx \exp\!\big(-\lambda_{\rm BL}A_2(r_{{C_0}O})\big),
\label{eq:PFA}
\end{equation}
where $A_2(r)$ is the effective blocking area under the nearest-building constraint
\begin{equation}
A_2(r)=4D^{2}\Bigl(\pi-\arccos\frac{D}{r}\Bigr)-2r^{2}\arcsin\frac{D}{r}+2D\sqrt{r^{2}-D^{2}}.
\label{eq:A2_area}
\end{equation}

\begin{IEEEproof}
Consider an OU at $\mathbf{U}$ and its nearest CIOS mounted on the building centered at $\mathbf{X}_0$, with $r_{{C_0}O} = |\mathbf{U} - \mathbf{X}_0|$. A building centered at $\mathbf{X}_i \neq \mathbf{X}_0$ would block the CIOS-OU link if its center lies inside the region obtained by taking the Minkowski sum of the segment $\overline {{{\bf{X}}_{\bf{0}}}{\bf{U}}}$ with a disk of radius $D$. The full blocking region, denoted $S(r)$, is bounded by the common external tangents to two circles: a circle of radius $D$ centered at $\mathbf{X}_0$ and a circle of radius $2D$ centered at ${\bf{U}}$. Its area is
\begin{equation}
		S(r) = 4\pi D^{2} + 3D\Bigl(\sqrt{r^{2}-D^{2}} - D\arccos\frac{D}{r}\Bigr).
		\label{eq:S_full}
\end{equation}
Because $\mathbf{X}_0$ is the nearest building center to the user, any other building center $\mathbf{X}_i$ must satisfy $\|\mathbf{X}_i - \mathbf{U}\| \ge r_{{C_0}O}$. This constraint removes the part of $S(r_{{C_0}O})$ that lies inside the disk of radius $r_{{C_0}O}$ centered at $\mathbf{U}$. The remaining effective blocking region (orange shaded area in Fig.~3) is $A_2(r_{{C_0}O})$ given in \eqref{eq:A2_area}. 
 
 Then, applying the void probability of the HPPP, we can yield \eqref{eq:PFA}.
\end{IEEEproof}

\emph{Lemma 5 (Conditional LoS Probability for the BS 0--CIOS 0 Link)}. Conditioned on $r_{{B_0}{C_0}}$, the conditional probability $\Pr({\cal B}\mid{\cal A},{\cal C})$ is approximated by
\begin{equation}
P_{f_B}(r_{{B_0}{C_0}})\approx \exp\!\big(-\lambda_{\rm BL}A_3(r_{{B_0}{C_0}})\big),
\end{equation}
where
\begin{equation}
A_3(r)=3D\Bigl(\sqrt{r^{2}-D^{2}}-D\arccos\frac{D}{r}\Bigr).
\label{eq:A3_area}
\end{equation}

 \begin{IEEEproof}
Under ${\cal A}$, the CIOS 0--OU segment is unblocked, which excludes a portion of potential blockers near that segment. Under ${\cal C}$, the BS and the OU must share an overlap on the serving facade, which further restricts the relative geometry and reduces the admissible blocking region for the BS 0--CIOS 0 segment. As a result, the effective blocking area for ${\cal B}$ depends on the overlap angle $\theta_i$ induced by ${\cal C}$ and can be written as $A_3(r,\theta_i)$, as depicted in Fig.~3 (the yellow shaded area).

An exact evaluation of $\Pr({\cal B}\mid{\cal A} ,{\cal C})$ would require retaining the overlap angle-dependent effective blocking area given by ${A_3}\left( {{r_{{B_0}{C_0}}},{\theta _i}} \right) = 3D\sqrt {r_{{B_0}{C_0}}^2 - {D^2}} - 3{D^2}\arccos \left( {\frac{D}{{{r_{{B_0}{C_0}}}}}} \!\right)- 4{D^2}\tan \left( {\frac{{{\theta _i}}}{2}} \right) + 2{D^2}{\theta _i}$, and then averaging $\exp(-\lambda_{\rm BL}A_3(r,\theta_i))$ over the distribution of $\theta_i$, which leads to an analytically cumbersome integral. To obtain a tractable expression, we adopt the approximation $\tan \left( {{{{\theta _i}} \mathord{\left/
 {\vphantom {{{\theta _i}} 2}} \right.
 \kern-\nulldelimiterspace} 2}} \right) \approx {{{\theta _i}} \mathord{\left/
 {\vphantom {{{\theta _i}} 2}} \right.
 \kern-\nulldelimiterspace} 2}$
which removes the explicit $\theta_i$-dependence in $A_3(r,\theta)$ and yields the closed-form effective area $A_3(r)$ in \eqref{eq:A3_area}. The void probability then gives
\begin{equation}
P_{f_B}(r_{{B_0}{C_0}})\approx \exp\!\big(-\lambda_{\rm BL}A_3(r_{{B_0}{C_0}})\big).
\end{equation}
\end{IEEEproof}
	
\emph{Lemma 6 (Visibility-Overlap Probability).}
Conditioned on $(r_{{B_0}{C_0}},r_{{C_0}O})$, the probability of ${\cal C}$ is
\begin{equation}
	{P_{{f_C}}}\left( {{r_{{B_0}{C_0}}},{r_{{C_0}O}}} \right) = \frac{{{\omega _B}\left( {{r_{{B_0}{C_0}}}} \right) + {\omega _U}\left( {{r_{{C_0}O}}} \right)}}{{2\pi }}.
\end{equation}

\begin{IEEEproof}
Condition ${\cal C}$ occurs iff the two visible arc intervals on the facade, with lengths $\omega_B$ and $\omega_U$, overlap. Since the relative orientation between BS 0 and the OU around the building is uniform over $[0,2\pi)$, the overlap probability equals the fraction of orientations that place one arc within a window of total length $\omega_B+\omega_U$, yielding $(\omega_B+\omega_U)/(2\pi)$.
\end{IEEEproof}

\emph{Corollary 2 (Successful Reflection Probability).}
Combining \eqref{eq:Pf_factor} with Lemmas 4--6, the conditional success probability of the serving reflection link is
\begin{equation}
	\begin{aligned}
		{P_f}(r_{{C_0}O},  r_{{B_0}O}, \varphi_O) &= \frac{1}{\pi} \left[ \arccos\left( \frac{D}{r_{{C_0}O}} \right) + \arccos\left( \frac{D}{r_{{B_0}{C_0}}} \right) \right] \\
		&\times \exp \left( { - {\lambda _{{\rm{BL}}}}\left( {{A_2}({r_{{C_0}O}}) + {A_3}({r_{{B_0}{C_0}}})} \right)} \right).
	\end{aligned}
\end{equation}

\emph{Corollary 3 (Reflected Interference Success Probability).} 
For an interfering BS $i$ $(i\neq 0)$, a building equipped with a CIOS can generate a specularly reflected interference path toward the OU. Conditioned on the BS--OU distance $r$ and the perpendicular offset $\rho$ of the reflecting building center to the BS--OU line, the two hops have equal length $d=\sqrt{\rho^2+r^2/4}$, and the corresponding success probability is
\begin{equation}
P_f^I(\rho,r)=\exp\!\big(-2\lambda_{\rm BL}S(d)\big),
\end{equation}
where $S(\cdot)$ is the blocking area in \eqref{eq:S_full}. Moreover, the offset $\rho_i$ follows
\begin{equation}
f_{\rho_i}(\rho)=\frac{4\lambda_{\rm BL}D}{\pi}\exp\!\left(-\frac{4\lambda_{\rm BL}D}{\pi}\rho\right).
\end{equation}

\begin{IEEEproof}
Under specular reflection, a reflected interference path exists when the perpendicular bisector of the BS $i$--OU segment intersects a building footprint, and a reflecting building at offset $\rho$ yields two equal-length hops $d=\sqrt{\rho^2+r^2/4}$. Each hop is unblocked with probability $\exp(-\lambda_{\rm BL}S(d))$ by the same blockage model, and assuming independent blockage on the two hops gives $P_f^I(\rho,r)=\exp(-2\lambda_{\rm BL}S(d))$. The PDF of $\rho_i$ follows from the void probability of the line-segment blockage model over a segment of length $\rho$, which yields $\Pr(\rho_i>\rho)=\exp(-4\lambda_{\rm BL}D\rho/\pi)$ and hence the stated exponential density.
\end{IEEEproof}

\vspace{-0.25cm}
\subsection{Transmission Link}
\vspace{-0.1cm}
For an IU, the association selects the nearest outdoor BS to the center of its host building. Owing to the nearest-building association, the serving BS--building-center segment is subject to the same nearest-building constraint as in the reflective case. Hence, the blockage characterization can be directly obtained by reusing the corresponding result, with the effective blocking area given by $A_2(\cdot)$.

\emph{Lemma 7 (LoS Probability for Serving Transmission Links).}
Conditioned on the BS--building-center distance $r_{{B_0}{L_0}}$, the probability that the serving transmission link is unblocked is
\begin{equation}
P_s(r_{{B_0}{L_0}})=\exp\!\big(-\lambda_{\rm BL}A_2(r_{{B_0}{L_0}})\big),
\label{eq:T_area}
\end{equation}
where $A_2(r)$ is the effective blocking area under the nearest-building constraint given in \eqref{eq:A2_area}.

\begin{IEEEproof}
Under the adopted association scheme, the serving BS--building segment follows the same nearest-building exclusion geometry as the serving BS--CIOS segment in the reflective case, which directly yields \eqref{eq:T_area}.
\end{IEEEproof}

\emph{Corollary 4 (LoS Probability for Interfering Transmission Links).}
Conditioned on an interfering distance $r$, the unblocked probability of an interfering transmission link follows the same segment-blockage model as \eqref{eq:S_full}, namely
\begin{equation}
P_s^I(r)=\exp\!\big(-\lambda_{\rm BL}S(r)\big).
\end{equation}

\vspace{-0.2cm}
\section{Conditional Coverage Probability for Outdoor and Indoor Users}
\vspace{-0.1cm}
The coverage probability is defined as
\begin{equation}
P_{\rm cov}(T)\triangleq \Pr(\mathrm{SIR}>T),
\end{equation}
where $T$ is the target threshold and $\mathrm{SIR}$ is defined as the ratio between the received signal power and the aggregate interference power. In this work, we focus on the interference-limited regime and thus neglect thermal noise in the subsequent analysis. We first derive the coverage probability for a typical OU by accounting for the serving link selection between the direct and CIOS-assisted reflection links. We then obtain the coverage probability for a typical IU served via the transmission link. 

\vspace{-0.5cm}
\subsection{Conditional Coverage Probability for Outdoor Users}
\vspace{-0.1cm}
A typical OU is served either by the direct BS 0--OU link or, when the direct link is blocked, by the BS 0--CIOS 0--OU reflection link. The interference consists of the aggregate direct interference from all other BSs and the aggregate reflected interference from CIOS-assisted paths. The following theorem gives the conditional OU coverage probability under the considered CIOS operating protocols.

\emph{Theorem 1 (Conditional Coverage Probability for an OU).}
Conditioned on $(r_{{B_0}O},r_{{C_0}O},\varphi_O)$, the OU coverage probability is provided at the bottom of this page, where $P_d(r_{{B_0}O})$ is the direct link success probability, $P_f(r_{{C_0}O},r_{{B_0}O},\varphi_O)$ is the reflection link success probability, and $P(\mathrm{SIR}_d>T\mid r_{{B_0}O})$ and $P(\mathrm{SIR}_r>T\mid r_{{B_0}O},r_{{C_0}O},\varphi_O)$ are the conditional SIR coverages of the direct and reflection links, respectively.

\begin{figure*}[hb]
	\hrulefill
		\begin{equation}
			\begin{split}
				P_O\left( \mathrm{SIR} > T \mid r_{{B_0}O}, r_{{C_0}O}, 
				\varphi_O \right) = & \; P_d(r_{{B_0}O}) \cdot P\left( \mathrm{SIR}_d > T \mid r_{{B_0}O} \right) \\
				& + \eta_O^{\rm TS} \cdot \left(1 - P_d(r_{{B_0}O})\right) \cdot P_f(r_{{C_0}O},  r_{{B_0}O}, \varphi_O) \cdot P\left( \mathrm{SIR}_r > T \mid r_{{B_0}O}, r_{{C_0}O}, \varphi_O \right),
			\end{split}	
			\label{eq:OU_coverage_decomposed}
		\end{equation}
\end{figure*}	

\begin{IEEEproof}
The OU coverage is determined by the successful connection probabilities and the corresponding conditional coverage probabilities of the direct and reflective links. Under TS, the reflective term is additionally weighted by $\eta_O^{\rm TS}$, yielding the result.
\end{IEEEproof}

\emph{Lemma 8 (Direct Link Coverage).}
Conditioned on the serving distance $r_{{B_0}O}$, the direct link SIR coverage is
\begin{equation}
P\!\left(\mathrm{SIR}_d>T\mid r_{{B_0}O}\right)
=\mathcal{L}_{I_d\mid r_{{B_0}O}}\!\left(\frac{T}{\ell_{{B_0}O}}\right)
\mathcal{L}_{I_r\mid r_{{B_0}O}}\!\left(\frac{\eta_O^{\rm ES}T}{\ell_{{B_0}O}}\right),
\label{eq:direct_link_coverage_laplace_opt}
\end{equation}
where $\ell_{{B_0}O}\triangleq \ell(r_{{B_0}O})$.

\begin{IEEEproof}
For the direct link, $\mathrm{SIR}_d = {g_{{B_0}O} \ell_{{B_0}O}}/{(I_d + {\eta_O^{\rm ES}}{I_r})}$,
with $g_{{B_0}O}\sim\exp(1)$. Conditioning on $(I_d,I_r)$ gives
\begin{equation}\nonumber
\Pr(\mathrm{SIR}_d>T\mid r_{{B_0}O},I_d,I_r)
=\exp\!\left(-\frac{T(I_d+\eta_O^{\rm ES}I_r)}{\ell_{{B_0}O}}\right).
\end{equation}
Taking expectation over $I_d$ and $I_r$ and using their independence yields
\begin{equation}\nonumber
P(\mathrm{SIR}_d>T\mid r_{{B_0}O})
=\mathbb{E}\!\left[e^{-\frac{T I_d}{\ell_{{B_0}O}}}\mid r_{{B_0}O}\right]
\mathbb{E}\!\left[e^{-\frac{\eta_O^{\rm ES}T I_r}{\ell_{{B_0}O}}}\mid r_{{B_0}O}\right],
\end{equation}
which is exactly \eqref{eq:direct_link_coverage_laplace_opt}.
\end{IEEEproof}

\emph{Lemma 9 (Reflection Link Coverage).}
Conditioned on ${R_O}\triangleq(r_{{C_0}O},r_{{B_0}O},\varphi_O)$, the reflection link SIR coverage is
\begin{equation}
	\begin{aligned}
	P&\left( \mathrm{SIR}_r > T \mid {R_O} \right) = \frac{1}{2}+ \\
	&\frac{1}{\pi} \int\limits_0^{ + \infty } \frac{\operatorname{Im}\!\left[ \mathcal{L}_{I_d|r_{{B_0}O}}{\left( {\frac{{j{t}}}{\eta_O^{\rm ES}}} \right)} \, \mathcal{L}_{I_r|r_{{B_0}O}}(jt) \, \mathcal{L}_{|h_{{B_0}{C_0}O}^{}|^2}\!{\left( {\frac{{ - j{t}}}{T}} \right)} \right]}{t}  dt,
	\label{eq:reflection_link_coverage_gilpelaez}
	\end{aligned}
\end{equation}
where $\mathcal{L}_{|h_{{B_0}{C_0}O}|^{2}}(s)$ denotes the Laplace transform of the serving cascaded channel power gain.

\begin{IEEEproof}
For the reflection link, ${\rm{SI}}{{\rm{R}}_r} = {{\eta _O^{{\rm{ES}}}|{h_{{B_0}{C_0}O}}{|^2}} \mathord{\left/
 {\vphantom {{\eta _O^{{\rm{ES}}}|{h_{{B_0}{C_0}O}}{|^2}} {\left( {{I_d} + \eta _O^{{\rm{ES}}}{I_r}} \right)}}} \right.
 \kern-\nulldelimiterspace} {\left( {{I_d} + \eta _O^{{\rm{ES}}}{I_r}} \right)}}$.
Let $X\triangleq I_d/\eta_O^{\rm ES}+I_r$. Then
\begin{equation}\nonumber
P(\mathrm{SIR}_r>T\mid R_O)
=\mathbb{E}_{|h|^{2}}\!\left[
F_X\!\left(\frac{|h|^{2}}{T}\right)\,\Big|\,R_O
\right].
\end{equation}
Using the Gil--Pelaez inversion theorem,
\begin{equation}\nonumber
F_X(x)=\frac{1}{2}-\frac{1}{\pi}\int_{0}^{\infty}\frac{\operatorname{Im}\!\left[e^{-jtx}\phi_X(t)\right]}{t}\,dt,
\end{equation}
where $\phi_X(t)=\mathbb{E}[e^{jtX}]$. Noting that
\begin{equation}\nonumber
{\phi _X}(t) = {{\cal L}_{{I_d}\mid {r_{{B_0}O}}}}\left( {{{ - jt} \mathord{\left/
 {\vphantom {{ - jt} {\eta _O^{{\rm{ES}}}}}} \right.
 \kern-\nulldelimiterspace} {\eta _O^{{\rm{ES}}}}}} \right){\mkern 1mu} {{\cal L}_{{I_r}\mid {r_{{B_0}O}}}}( - jt),
\end{equation}
and exchanging the expectation with the integral yields \eqref{eq:reflection_link_coverage_gilpelaez} after recognizing $\mathbb{E}[e^{\frac{jt}{T}|h|^{2}}]=\mathcal{L}_{|h|^{2}}\!\left(\frac{-jt}{T}\right)$.
\end{IEEEproof}

Lemmas~8 and~9 show $P(\mathrm{SIR}_d>T\mid r_{{B_0}O})$ and $P(\mathrm{SIR}_r>T\mid r_{{B_0}O},r_{{C_0}O},\varphi_O)$ in terms of the conditional Laplace transforms of the direct and reflected interference and the Laplace transform of the serving cascaded channel power gain. Next, we derive the Laplace transforms of $I_d$, $I_r$, and $|h|^{2}$.

\emph{Lemma 10 (Laplace Transform of Direct Interference).}
Conditioned on the serving distance $r_{{B_0}O}$, the Laplace transform of the aggregate direct interference from all non-serving BSs is given by
\begin{equation}
	\mathcal{L}_{I_d|{r_{{B_0}O}}}(s) = \exp\!\left( -2\pi \lambda _{\rm{BS}}^{(0)}\int_{{r_{{B_0}O}}}^{+\infty} \frac{s \, P_d^I(r) \, \ell(r)}{1 + s \, \ell(r)} \, r \, dr \right),
	\label{eq:laplace_direct_interference}
\end{equation}

\begin{IEEEproof}
The aggregate direct interference is
\begin{equation}\nonumber
I_d=\sum_{i\in\Phi_{\rm BS}\setminus\{0\}} \zeta_i\, g_{{B_i}O}\,\ell(r_i),
\end{equation}
where $r_i=\|\mathbf{B}_i-\mathbf{U}\|$, $g_{{B_i}O}\sim\exp(1)$, and $\zeta_i\in\{0,1\}$ indicates whether the BS$_i$--OU link is unblocked with $\Pr(\zeta_i=1\mid r_i)=P_d^I(r_i)$.

Conditioned on $r_{{B_0}O}$ and using independence across interferers,
\begin{equation}\nonumber
\mathcal{L}_{I_d\mid r_{{B_0}O}}(s)
=\mathbb{E}_{\Phi_{\rm BS}}\!\left[\prod_{i\in\Phi_{\rm BS}\setminus\{0\}}
\mathbb{E}_{\zeta_i,g}\!\left(e^{-s\zeta_i g\ell(r_i)}\right)\,\Big|\,r_{{B_0}O}\right].
\end{equation}
Since $g\sim\exp(1)$, $\mathbb{E}_{\zeta_i,g}\!\left(e^{-s\zeta_i g\ell(r_i)}\right)
=(1-P_d^I(r_i))+\frac{P_d^I(r_i)}{1+s\ell(r_i)}$.
Approximating the outdoor BS process by an equivalent PPP of density $\lambda _{\rm{BS}}^{(0)}$ and applying its probability generating functional (PGFL) $\mathbb{E}\left[ {{\Pi _{x \in \Psi }}f\left( x \right)} \right] = \exp \left( { - \lambda \int_{{\mathbb{R}^2}} {\left( {1 - f\left( x \right)} \right)dx} } \right)$ over $\mathbb{R}^2\setminus B(\mathbf{U},r_{{B_0}O})$ yields \eqref{eq:laplace_direct_interference}.
\end{IEEEproof}

For the reflected interference, we consider the default specular-reflection state without phase alignment. A reflected interference path from an interfering BS exists whenever a building equipped with a CIOS satisfies the specular geometry. For tractability, we adopt the center-based approximation that the reflecting building center lies on the perpendicular bisector of the BS--OU segment. For a given BS--OU distance $r$ and perpendicular offset $\rho$ of the reflecting building center to the BS--OU line, the two hops have equal length $d=\sqrt{\rho^2+r^2/4}$, and the corresponding reflected-path success probability is $P_f^I(\rho,r)$ in Corollary~3.

\emph{Lemma 11 (Laplace Transform of Reflected Interference).}
Conditioned on $r_{{B_0}O}$, the Laplace transform of the aggregate reflected interference from non-serving CIOSs is
\begin{equation}
\mathcal{L}_{I_r\!\mid\! r_{{B_0}O}}\!(\!s\!)\!
\!=\!\exp\!\left(\!
-2\pi\lambda _{\rm{BS}}^{(0)}\!\!\!\!\!\int\limits_{r_{{B_0}O}}^{\infty}
\!\!\left[\int\limits_{0}^{\infty}
\!\!\!\!\!\frac{s\,a(\rho,r)}{1+s\,a(\rho,r)}\,
\!\!\!P_f^I(\rho,r)\, \!\!f_{\rho_i}(\!\rho\!)\, d\rho\!
\right]\! r\,\!dr\!\!
\right)
\label{eq:laplace_reflected_interference_opt}
\end{equation}
where
\begin{equation}
a(\rho,r)\triangleq N_{{\rm eff},O}'(\rho,r)\,d^{-\alpha},
\end{equation}
and
\begin{equation}
N_{{\rm eff},O}'(\rho,r)
=\frac{\eta_O^{\rm MS}N}{\pi}
\left(
\arccos\!\frac{D}{d}
-\arctan\!\frac{r}{2\rho}
\right).
\end{equation}

\begin{IEEEproof}
Please refer to Appendix B.
\end{IEEEproof}

\emph{Lemma 12 (Laplace Transform of the Serving Cascaded Reflection Gain).} Conditioned on $R_O=(r_{{C_0}O},r_{{B_0}O},\varphi_O)$, the Laplace transform of the serving cascaded-channel power gain $|h_{{B_0}{C_0}O}|^2$ is
	\begin{equation}
		\begin{aligned}
			&\mathcal{L}_{|h_{{B_0}{C_0}O}|^2|R_O}\!(s)\!
			\!= \!\!\!\!\int_0^{{\frac{{\min ({\omega _U},{\omega _B})}}{\pi }\eta _O^{{\rm{MS}}}N}} \!\!\!\!\!\!\!\!\!\!\!\!\!\!\!\!\!\!\!\!\!\!\!\!\!\frac{1}{{\sqrt {1 + 2s(1 - \frac{{{\pi ^2}}}{{16}})n\ell ({r_{{B_0}{C_0}}})\ell ({r_{{C_0}O}})} }}\!\!\\
			&\quad\cdot \exp \left( { - \frac{{s{\pi ^2}{n^2}{\mkern 1mu} \ell ({r_{{B_0}{C_0}}})\ell ({r_{{C_0}O}})}}{{16 + 2s(16 - {\pi ^2})n{\mkern 1mu} {\mkern 1mu} \ell ({r_{{B_0}{C_0}}})\ell ({r_{{C_0}O}})}}} \right){\mkern 1mu} {f_{{N_{{\rm{eff}},O}}}}(n){\mkern 1mu} dn.
			\label{eq:laplace_desired_reflection}
		\end{aligned}
	\end{equation}

\begin{IEEEproof}
	Since the cascade channel $|h_{{B_0}{C_0}O}|$ can be approximated by a Gaussian random variable as shown in \eqref{eq:signal_gaussion}. Equivalently, $|h_{{B_0}{C_0}O}|^2$ can be expressed as
	\begin{equation}
	|h_{{B_0}{C_0}O}|^2 = Z \cdot \Bigl(1 - \frac{\pi^2}{16}\Bigr) N_{\mathrm{eff},O} \, {\ell ({r_{{B_0}{C_0}}})\ell ({r_{{C_0}O}})},
	\end{equation}
	where $Z$ follows a non-central chi-squared distribution with one degree of freedom and non‑centrality parameter $\lambda = \frac{\mu^2}{\sigma^2} = \frac{N_{\mathrm{eff},O} \pi^2}{16 - \pi^2}$.

	The Laplace transform of $Z$ is known to be
	\begin{equation}
	\mathbb{E}\bigl[e^{-s Z}\bigr] = \frac{1}{\sqrt{1+2s}} \exp\!\Bigl( -\frac{\lambda s}{1+2s} \Bigr).
	\end{equation}
	Replacing $s$ by $s \bigl(1 - \frac{\pi^2}{16}\bigr) N_{\mathrm{eff},O} \, {\ell ({r_{{B_0}{C_0}}})\ell ({r_{{C_0}O}})}$ and averaging over the distribution of \(N_{\mathrm{eff},O}\) yields precisely \eqref{eq:laplace_desired_reflection}.
\end{IEEEproof}

\vspace{-0.3cm}
\subsection{Conditional Coverage Probability for Indoor Users}

A typical IU is served only through the CIOS transmission link of its host building, since no indoor transmitters are deployed. The aggregate interference is contributed by transmission links from non-serving BSs via other buildings equipped with CIOSs. 

\emph{Theorem 2 (Conditional Coverage Probability for an IU).}
Conditioned on $(r_{{B_0}{L_0}},r_{{L_0}I})$, the IU coverage probability is
\begin{equation}
P_I\!\left(\!\mathrm{SIR}\!>\!T\!\mid\! r_{{B_0}{L_0}},r_{{L_0}I}\!\right)
\!=\! \eta_I^{\rm TS} P_s(r_{{B_0}{L_0}})\,\!
P\!\left(\!\mathrm{SIR}_t\!>\!T\!\mid\! r_{{B_0}{L_0}},r_{{L_0}I}\!\right),
\label{eq:IU_coverage_decomposed}
\end{equation}
where $P_s(\cdot)$ is given in Lemma~7 and $P(\mathrm{SIR}_t>T\mid\cdot)$ is characterized below.
\begin{IEEEproof}
The IU coverage is determined by the successful connection probability of the serving transmission link and the corresponding conditional coverage probability. Under TS, the protocol effect is incorporated through $\eta_I^{\rm TS}$, which gives the stated expression.
\end{IEEEproof}

\emph{Lemma 13 (Transmission Link Coverage).}
For the transmission link, $\mathrm{SIR}_t=\eta_I^{\rm ES} |h_{{B_0}{C_0}I}|^2/I_t$. Conditioned on $(r_{{B_0}{L_0}},r_{{L_0}I})$, the coverage probability is
\begin{equation}
	\begin{aligned}
	P&\left( \mathrm{SIR}_t > T \mid {r_{{B_0}{L_0}}}, {r_{{L_0}I}} \right) = \frac{1}{2} +\\ &\frac{1}{\pi} \int_0^{+\infty} \frac{\operatorname{Im}\!\left[ \mathcal{L}_{I_t|{r_{{B_0}{L_0}}}, {r_{{L_0}I}}}(jt) \, \mathcal{L}_{|h_{{B_0}{C_0}I}|^2|{R_I}}\!\left( { - \frac{{j{t}}}{T}} \right) \right]}{t}  dt,
	\label{eq:transmission_link_coverage_gilpelaez}
	\end{aligned}
\end{equation}
where $\mathcal{L}_{I_t\mid r_{{B_0}{L_0}},r_{{L_0}I}}(\cdot)$ is the Laplace transform of the aggregate transmission interference and $\mathcal{L}_{|h_{{B_0}{C_0}I}|^2\mid R_I}(\cdot)$ is the Laplace transform of the serving transmission-channel power gain.

\begin{IEEEproof}
The proof follows the same Gil--Pelaez inversion argument as in Lemma~9. Conditioned on $(r_{{B_0}{L_0}},r_{{L_0}I})$, the transmission-link coverage is obtained by applying the inversion theorem to $I_t/\eta_I^{\rm ES}$, whose characteristic function is equivalently expressed through $\mathcal{L}_{I_t|r_{{B_0}{L_0}},r_{{L_0}I}}(\cdot)$. Taking the expectation over $|h_{{B_0}{C_0}I}|^2$ then gives \eqref{eq:transmission_link_coverage_gilpelaez}.
\end{IEEEproof}

\emph{Lemma 14 (Laplace Transform of Transmission Interference).}
Conditioned on $(r_{{B_0}{L_0}},r_{{L_0}I})$, the Laplace transform of the aggregate transmission interference from all non-serving BSs is
\begin{equation}
	\begin{aligned}
	{{\cal L}_{{I_t}|{r_{{B_0}{L_0}}},{r_{{L_0}I}}}}&(s) = \exp \left( { - \lambda _{\rm{BS}}^{(0)}} \right. \cdot \\
	&\int_0^{2\pi }\int_{{r_{{B_0}{L_0}}}}^{ + \infty } 
	{\frac{s \eta_I^{\rm ES} {N_{\text{eff},I}'}(r){P_s^I}(r){\ell _I}(r,\varphi)}{1 + s \eta_I^{\rm ES}{N_{\text{eff},I}'}(r){\ell _I}(r,\varphi )}}rdrd\varphi
	),
	\label{eq:laplace_transmission_interference}
	\end{aligned}
\end{equation}
where 
	\begin{equation}
		{N_{\text{eff},I}'}(r) = \frac{\eta_I^{\rm MS} N}{\pi} \arccos\left( \frac{D}{r} \right),
		\label{eq:NeffI'}
	\end{equation}

	\begin{equation}
		{\ell _I}(r,\varphi ) = {\sqrt {{r^2} + r_{{L_0}I}^2 - 2r{r_{{L_0}I}}\cos \varphi } ^{ - \alpha }}.
	\end{equation}

\begin{IEEEproof}
Please refer to Appendix C.
\end{IEEEproof}

\emph{Lemma 15 (Laplace Transform of the Serving Transmission Channel).} Conditioned on $R_I\triangleq(r_{{L_0}I},r_{{B_0}{L_0}},\varphi_I)$, the Laplace transform of the serving transmission channel power gain is
\begin{equation}
	\mathcal{L}_{|h_{{B_0}{C_0}I}|^2\!|\!{R_I}}\!(\!s\!)
	\!=\!\! \frac{\exp\!\Bigl( -\dfrac{s \eta_I^2 \pi N_{\mathrm{eff},I}^2(r_{{B_0}{L_0}}) \ell(r_{{B_0}I}) }
		{4 + 2 (4 - \pi) s \eta_I^{\rm ES} N_{\mathrm{eff},I}(r_{{B_0}{L_0}}) \ell(r_{{B_0}I}) } \Bigr)}
	{\sqrt{1 + s \dfrac{(4 - \pi)}{2} \eta_I^{\rm ES} N_{\mathrm{eff},I}(r_{{B_0}{L_0}}) \ell(r_{{B_0}I}) }}.
	\label{eq:laplace_desired_transmission}
\end{equation}

\begin{IEEEproof}
	Following the same Gaussian approximation as in Lemma 12, the channel gain can be expressed as
	\begin{equation}
	|h_{{B_0}{C_0}I}|^2 = Z \cdot \frac{(4 - \pi) N_{\mathrm{eff},I} \ell(r_{{B_0}I})}{4},
	\end{equation}
	where $Z$ follows a non-central chi-squared distribution with one degree of freedom. Similarly, its non-centrality parameter $\lambda =\frac{N_{\mathrm{eff},I} \pi}{4 - \pi}$, and substituting $s\eta_I^{\rm ES} \frac{(4-\pi)}{4} N_{\mathrm{eff},I} \ell(r_{{B_0}I})$ in the Laplace transform yields the result in \eqref{eq:laplace_desired_transmission}.
\end{IEEEproof}

\begin{figure*}[!t]
	\subfigure[OUs]{\includegraphics[width=0.49\textwidth]{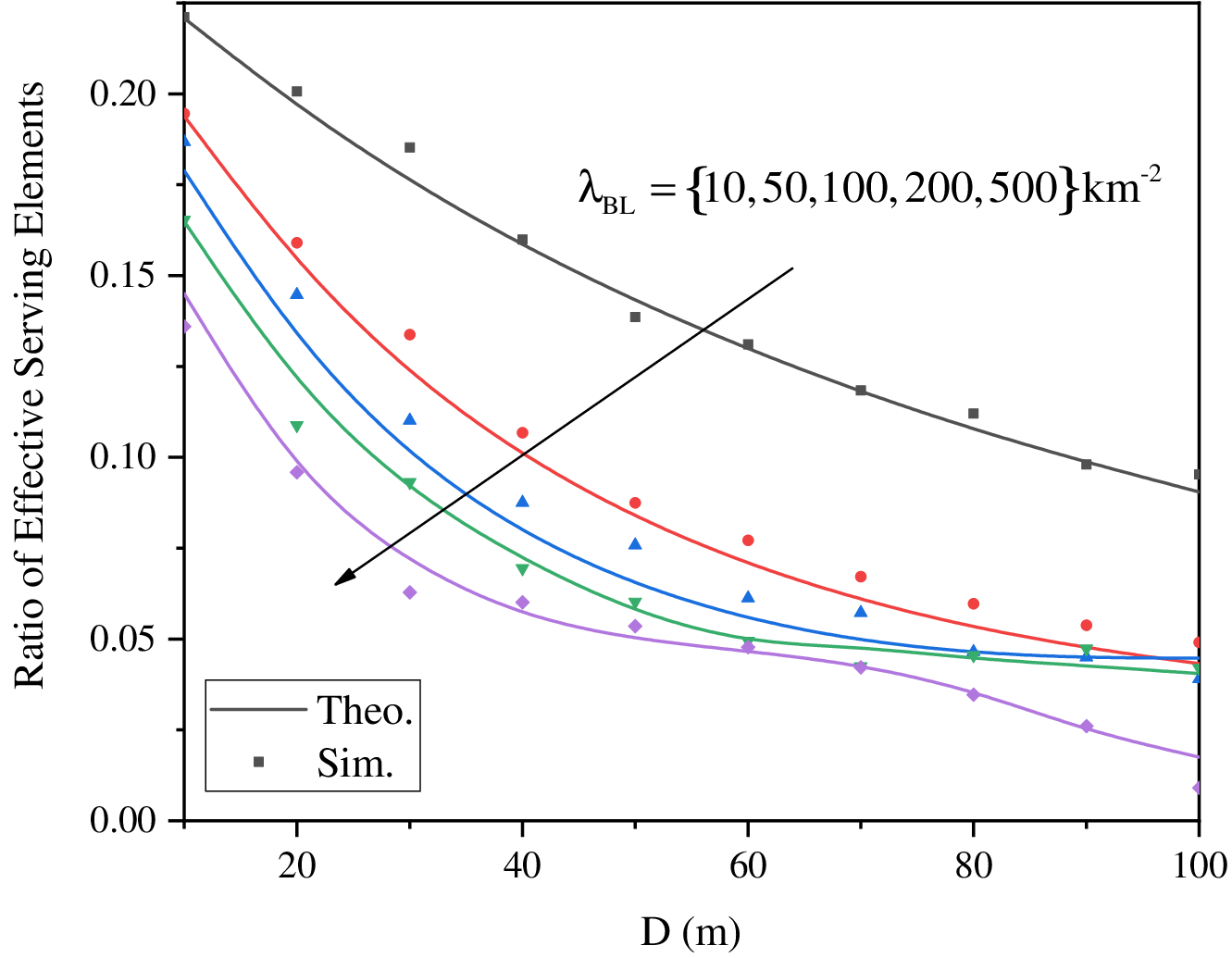}}
	\subfigure[IUs]{\includegraphics[width=0.48\textwidth]{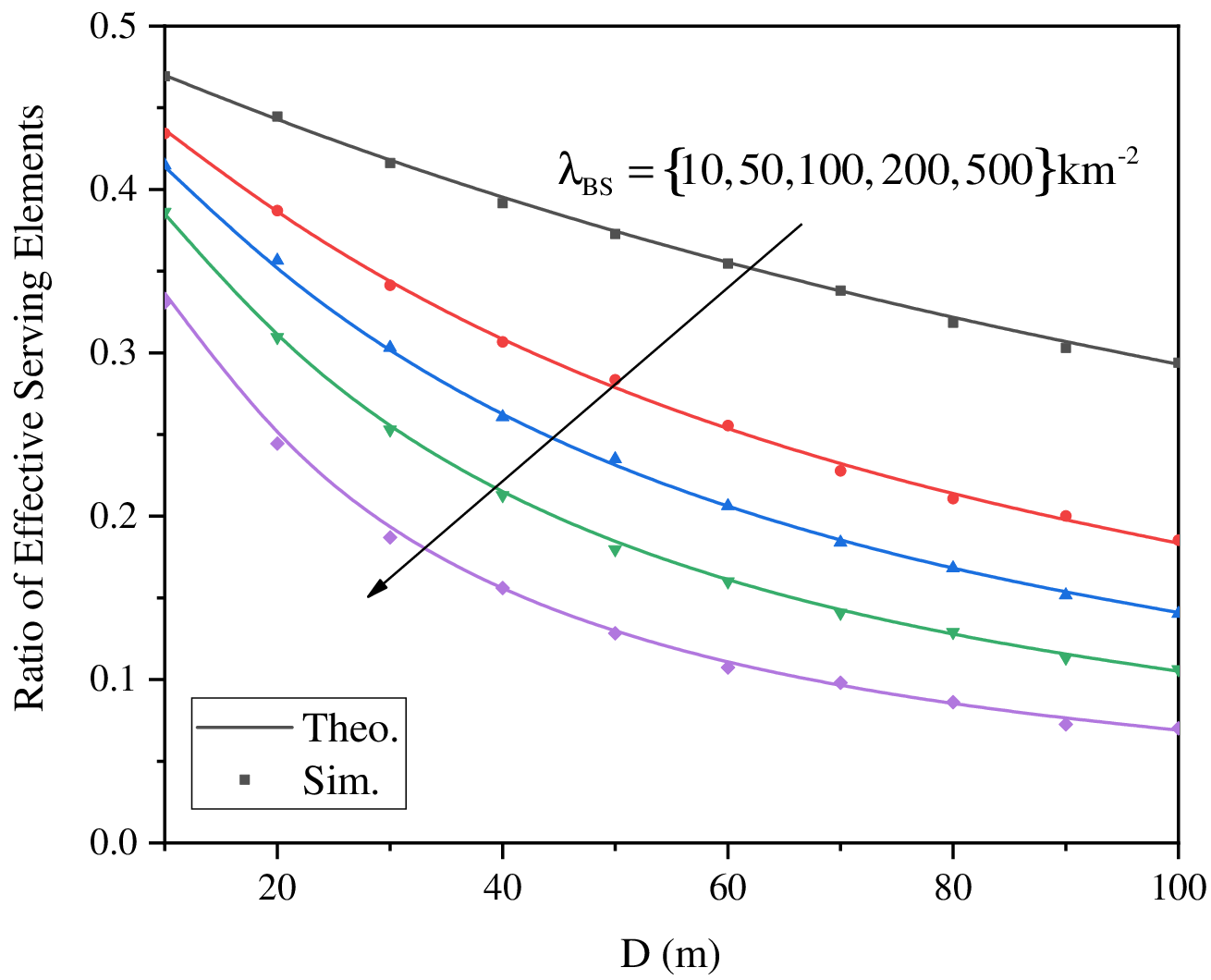}}
	\vspace{-0.3cm}
	\caption{Ratio of effective serving elements versus building radius $D$.}
	\vspace{-0.2cm}
\end{figure*}

\begin{figure*}[!t]
	\subfigure[Direct Link]{\includegraphics[width=0.33\textwidth]{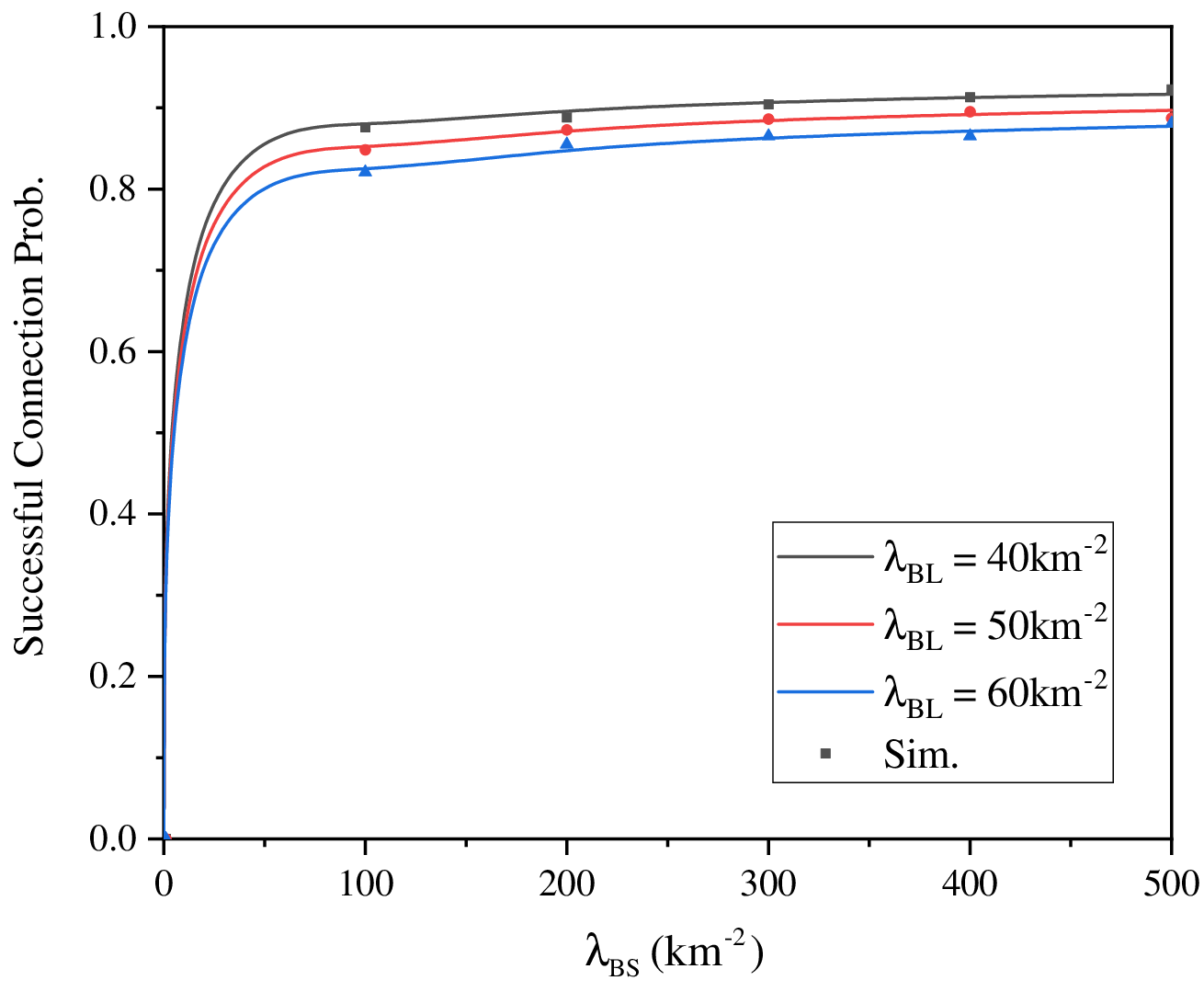}}
	\subfigure[Reflection Link]{\includegraphics[width=0.33\textwidth]{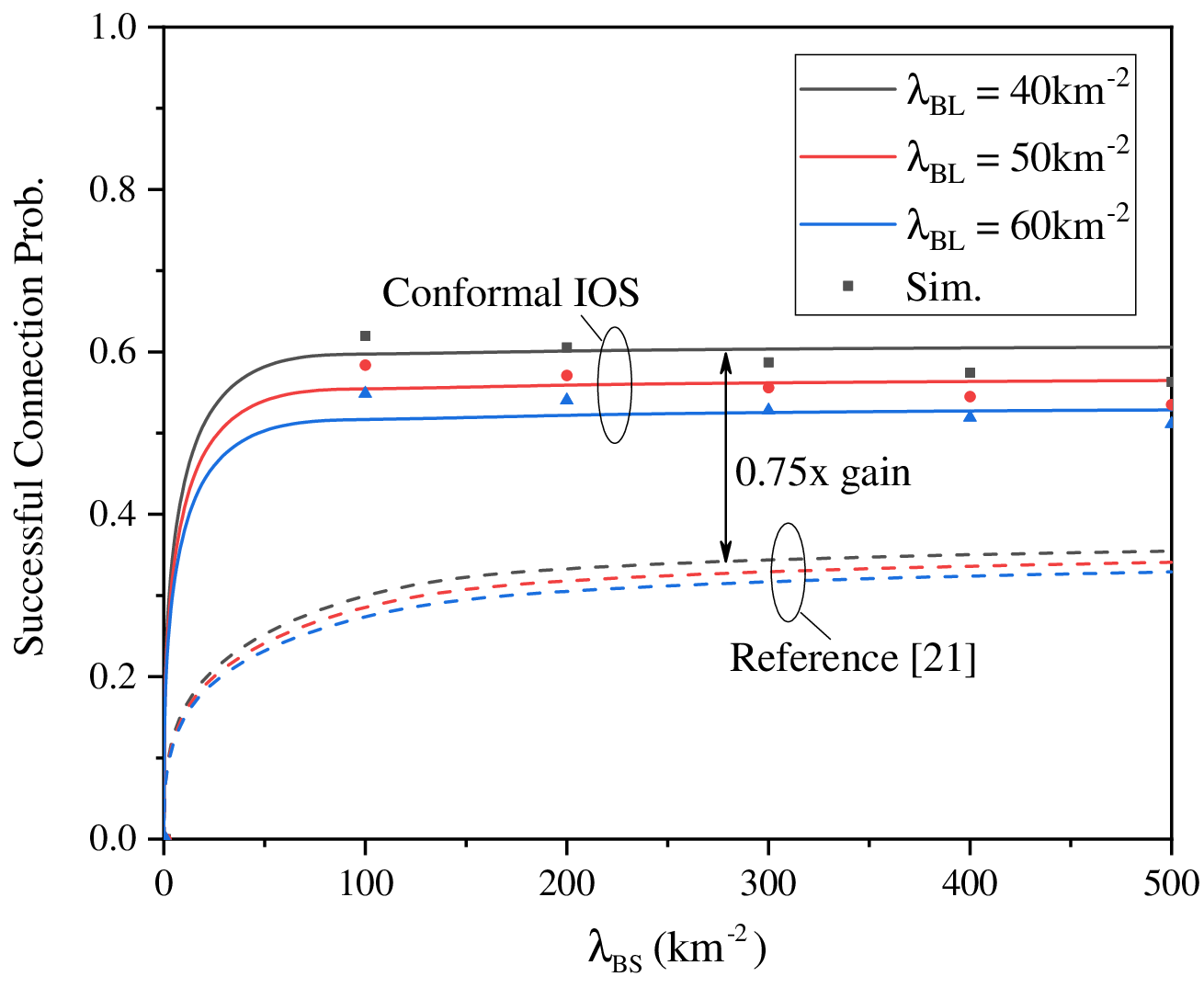}}
	\subfigure[Transmission Link]{\includegraphics[width=0.33\textwidth]{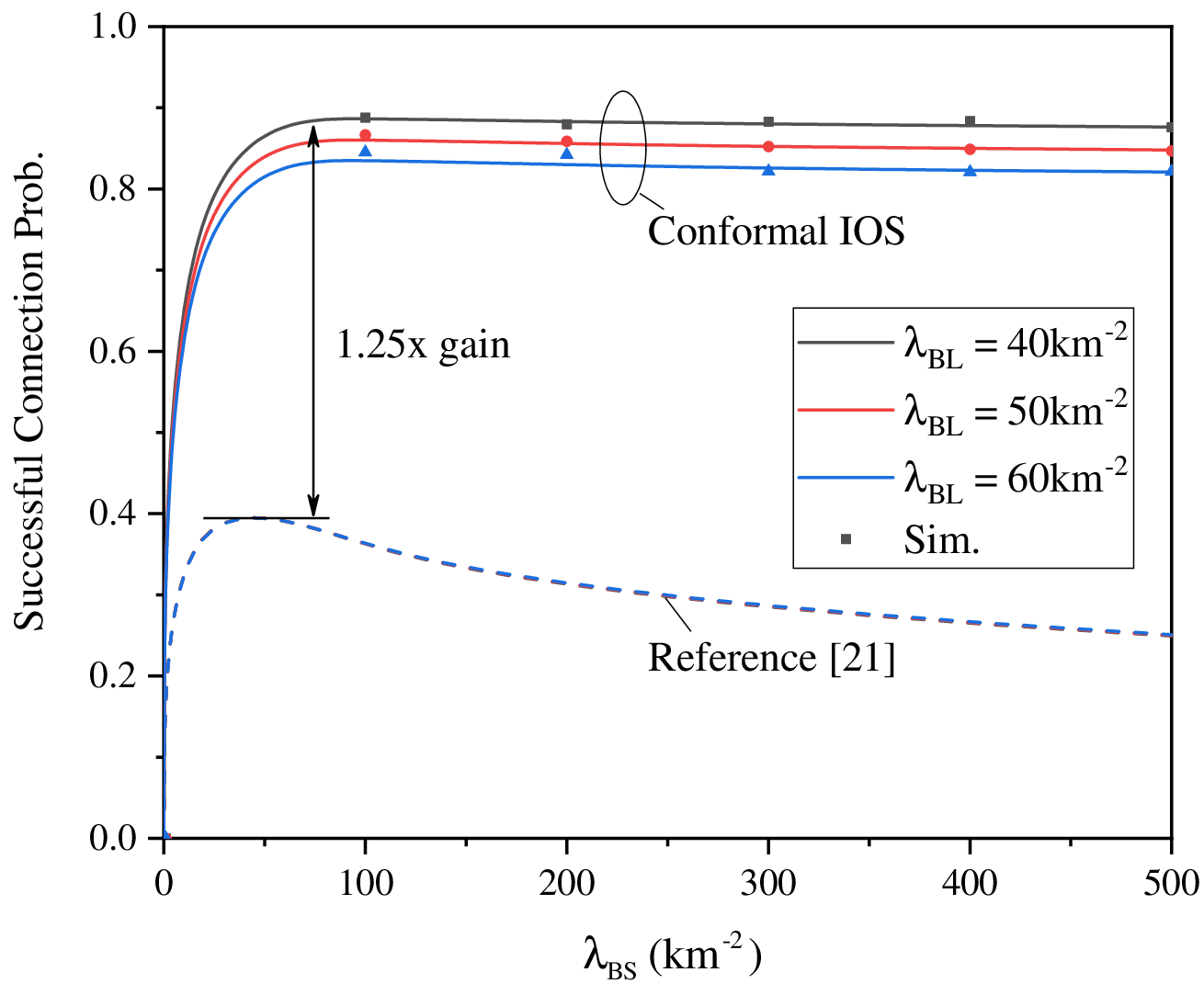}}
	\vspace{-0.5cm}
	\caption{Successful connection probability versus BS densities $\lambda_{\rm BS}$ for different building densities $\lambda_{\rm BL}$.}
	\vspace{-0.2cm}
\end{figure*}

\section{Overall Coverage Probability}

\emph{Theorem 3 (Overall Coverage Probability).}
The probability that a randomly selected user achieves $\mathrm{SIR}>T$ is
\begin{equation}
P_c(T)= w_I\,P_I(T)+ w_O\,P_O(T),
\label{eq:overall_coverage}
\end{equation}
where $P_I(T)$ and $P_O(T)$ denote the spatially averaged coverage probabilities for IUs and OUs, and $(w_I,w_O)$ are the corresponding user-type selection probabilities.

The weights follow from the effective outdoor-area thinning and are given by
\begin{equation}
w_I=\frac{\lambda_{\rm IU}}{\lambda_{\rm IU}+\lambda_{\rm OU}e^{-\lambda_{\rm BL}\pi D^2}},
\qquad
w_O=1-w_I.
\label{eq:weights}
\end{equation}
The spatial averages are obtained by integrating the conditional coverage expressions over the associated geometric random variables:
\begin{equation}
	\begin{aligned}
		&P_O(T) \!=\! \int_{0}^{2\pi} \!\!\int_{D}^{\infty} \!\!\int_{0}^{\infty}\!\! \!P_O\left( \mathrm{SIR} > T \mid r_{{B_0}O}, r_{{C_0}O}, 
		\varphi_O \right)\times \\
		& \frac{ f_{r_{{B_0}O}}(r_{{B_0}O}) \, f_{r_{{C_0}O}}(r_{{C_0}O}) \, f_{N_{\mathrm{eff},O}}(n) }{ 2\pi }  \, dr_{{B_0}O} \, dr_{{C_0}O} \, d\varphi_O, 
		\label{eq:avg_PO} \\[6pt]
	\end{aligned}
\end{equation}

\begin{equation}
	\begin{aligned}
	P_I(T) &= \int_{0}^{2\pi} \!\int_{D}^{\infty}\! \int_{0}^{D}\! P_I\left( \mathrm{SIR} > T \mid r_{{L_0}I},r_{{B_0}{L_0}},\varphi _I \right) \\
	&\quad \times \frac{ f_{r_{{L_0}I}}(r_{{L_0}I}) \, f_{r_{{B_0}{L_0}}}(r_{{B_0}{L_0}}) }{ 2\pi } \, dr_{{L_0}I} \, dr_{{B_0}{L_0}} \, d\varphi_I. 
	\label{eq:avg_PI}
	\end{aligned}
\end{equation}

\begin{IEEEproof}
Since the conditional indoor and outdoor coverage probabilities have been obtained, the overall coverage follows by weighting them according to the user densities and averaging over the corresponding distance distributions.
\end{IEEEproof}

\section{Numerical Results}
\vspace{-0.1cm}
In this section, we present the numerical results for the effective serving elements ratio for both the OU and IU, the probability of successful connection, and the coverage probability, all of which have been derived in the preceding sections. Unless stated otherwise, the system parameters are set to $\alpha  = 2.5$, $N=100$, $T=0$ dB, $D=20$ m, $\eta _I^p = 0.5$, and ${\lambda _{{\rm{IU}}}}/{\lambda _{{\rm{OU}}}} = 1$. To validate the theoretical analysis, Monte Carlo simulations are conducted, demonstrating an agreement between the analytical results and the simulation outcomes.			

\begin{figure*}[!t]
	\subfigure[$D=10$m]{\includegraphics[width=0.48\textwidth]{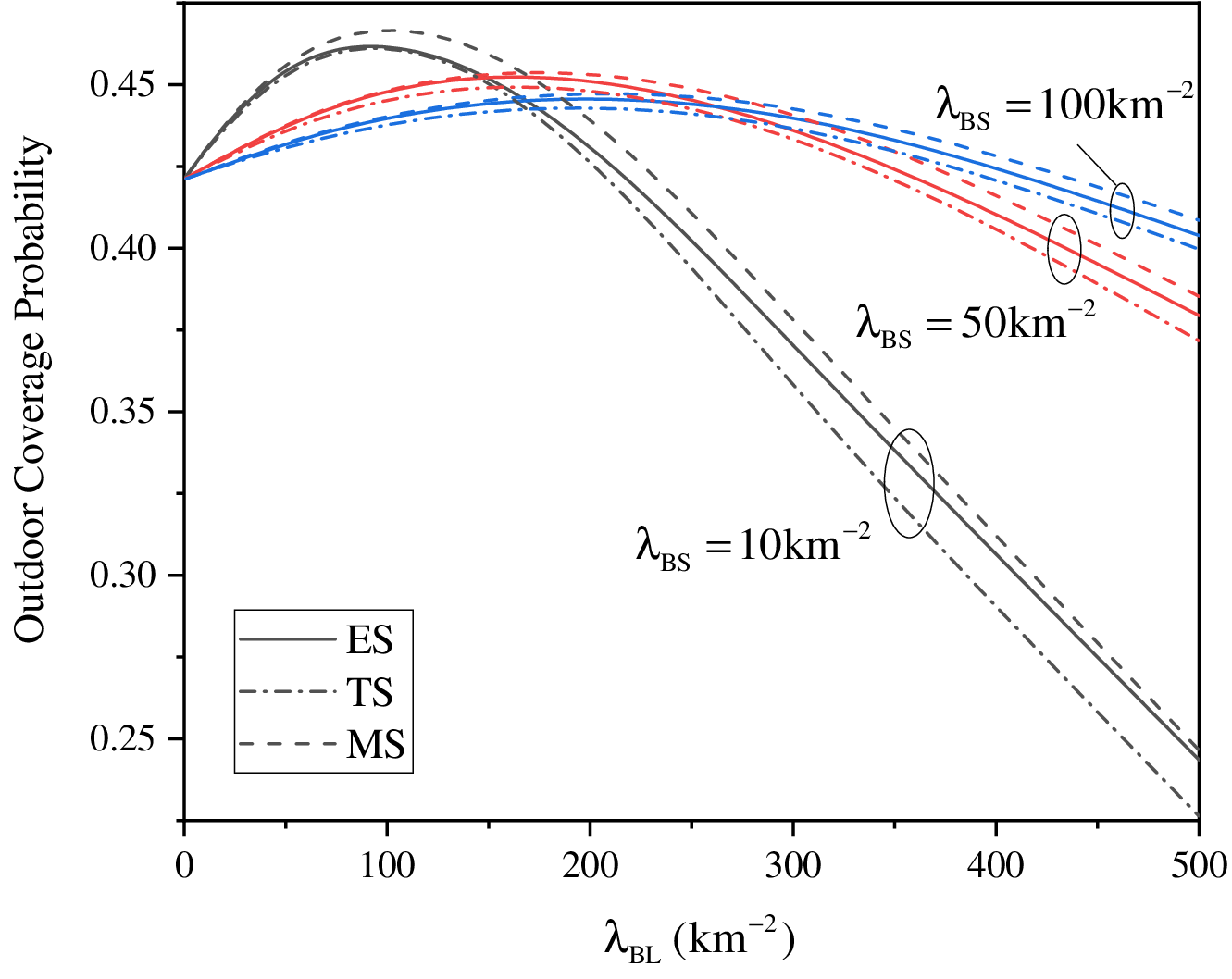}}
	\subfigure[$D=20$m]{\includegraphics[width=0.48\textwidth]{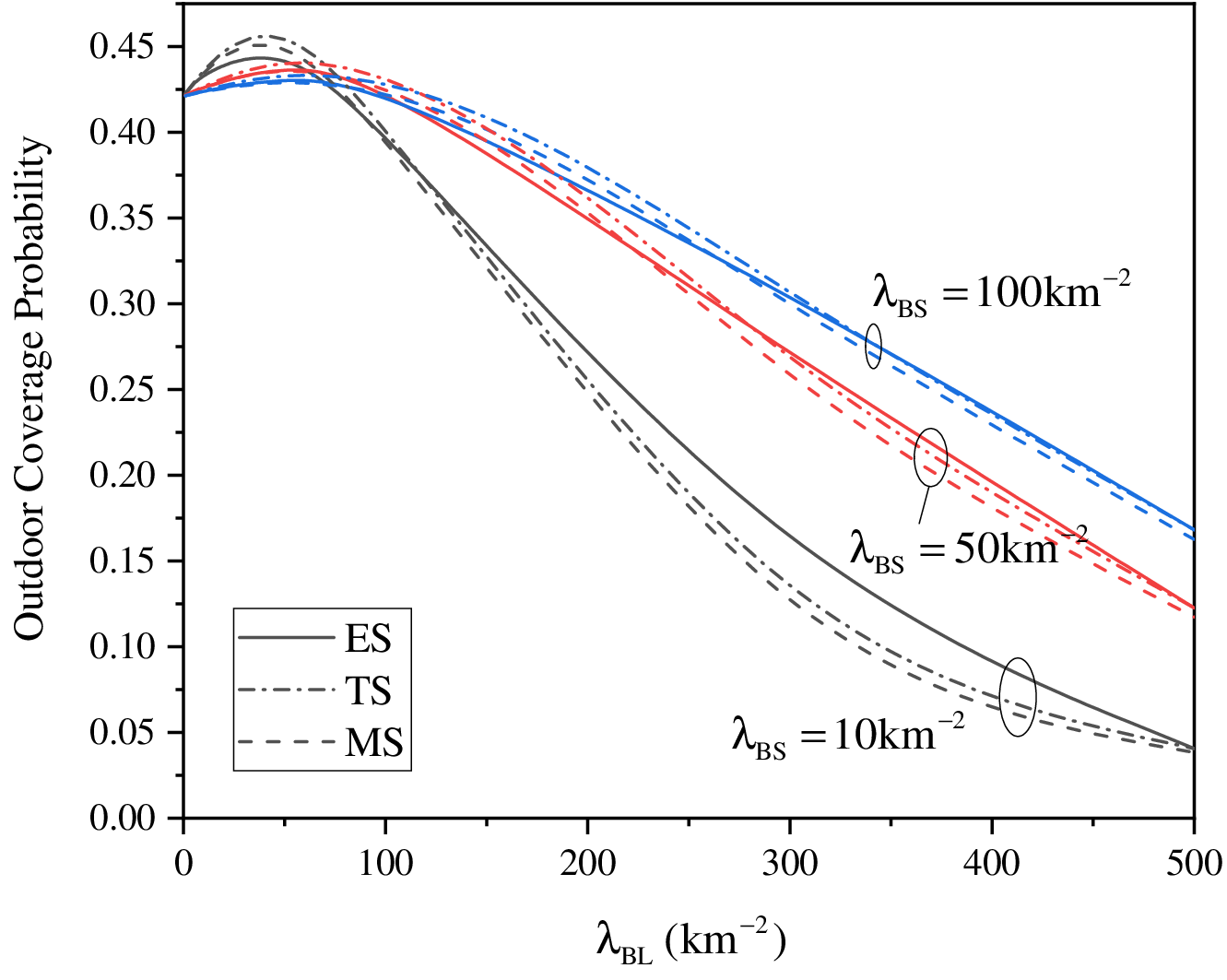}}
	\vspace{-0.3cm}
	\caption{Coverage probability for OUs versus building density $\lambda_{\rm BL}$ for different BS densities $\lambda_{\rm BS}$ under three operating protocols.}
	\vspace{-0.2cm}
\end{figure*}

\begin{figure*}[!t]
	\subfigure[$D=10$m]{\includegraphics[width=0.48\textwidth]{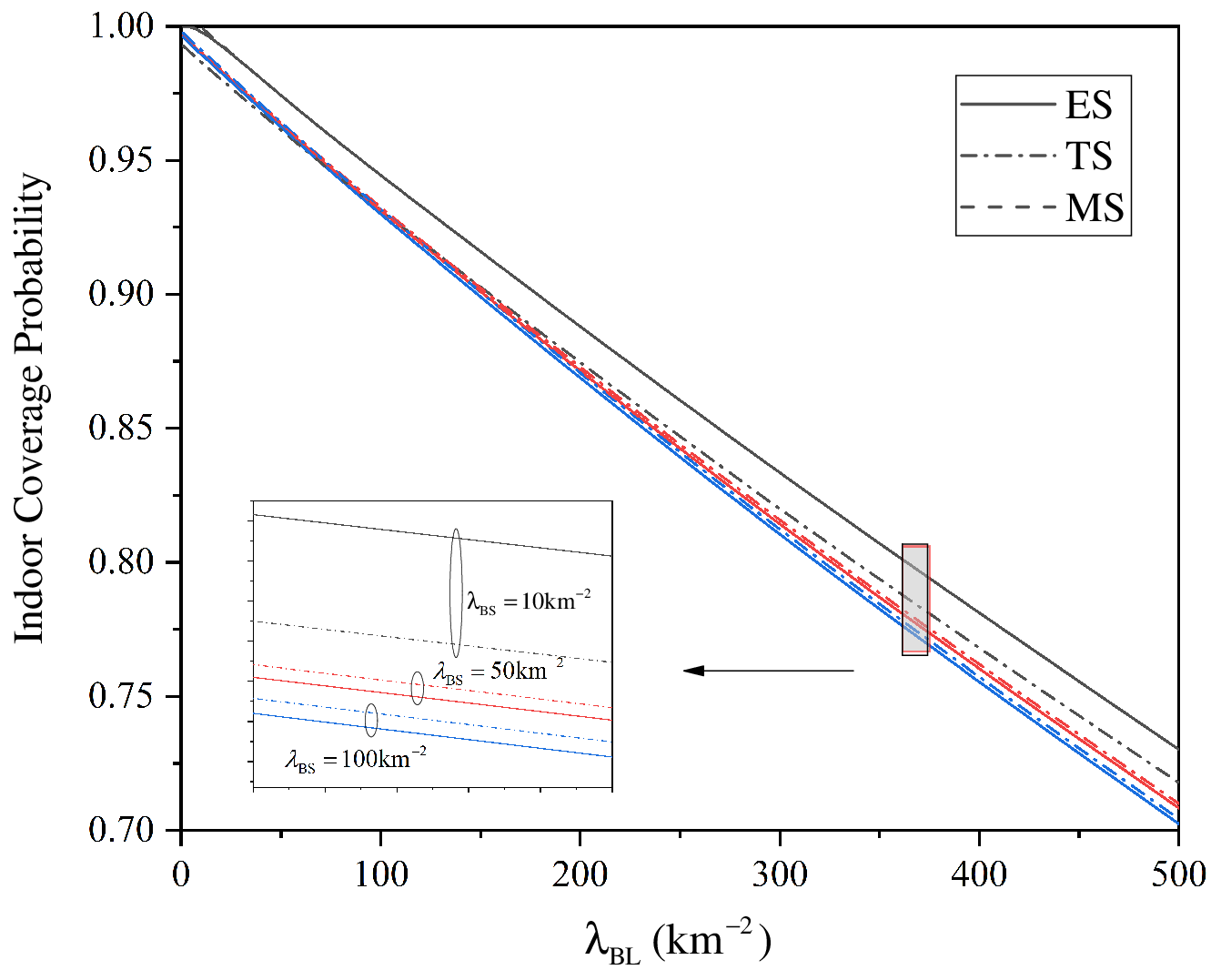}}
	\subfigure[$D=20$m]{\includegraphics[width=0.48\textwidth]{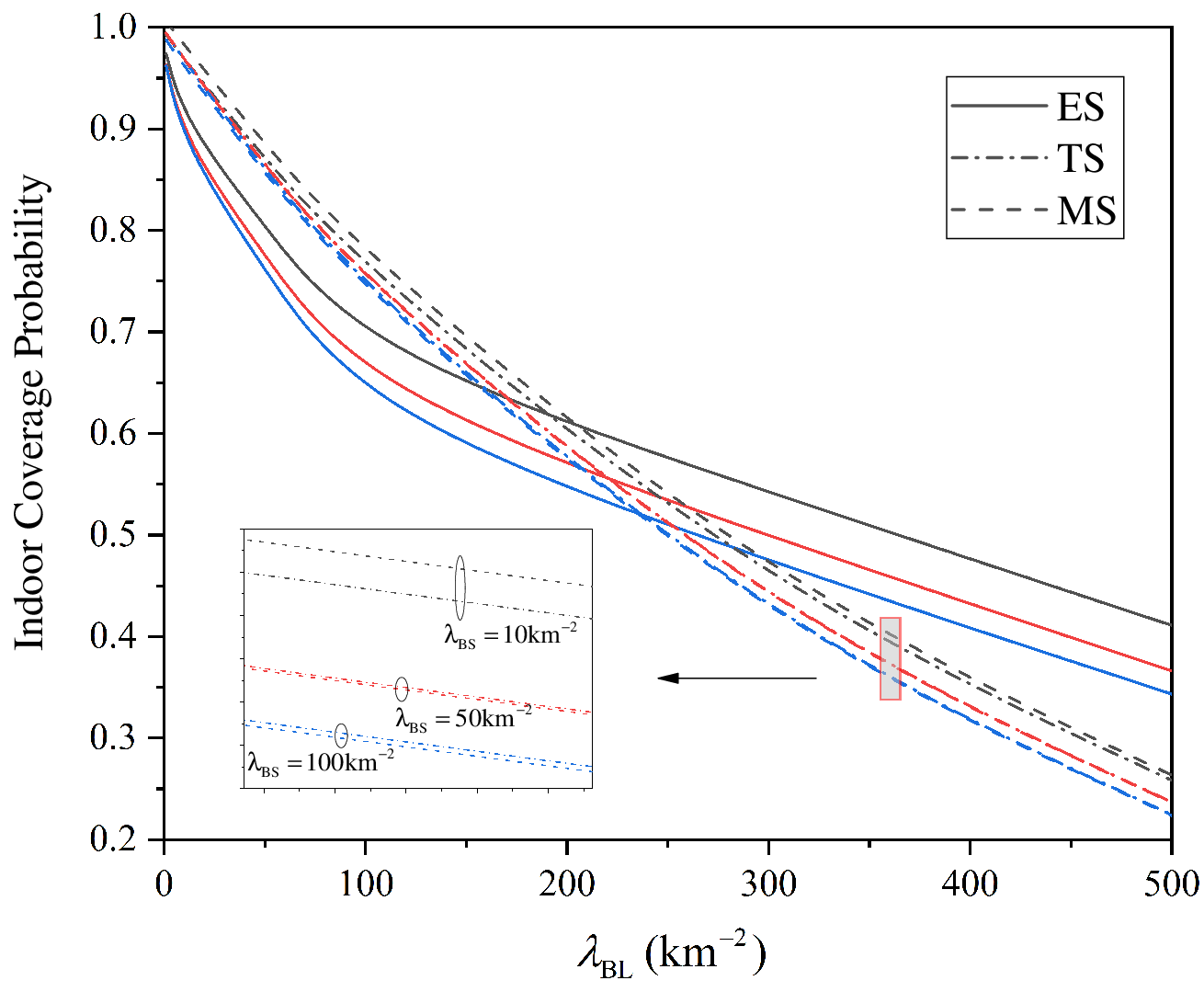}}
	\vspace{-0.3cm}
	\caption{Coverage probability for IUs versus building density $\lambda_{\rm BL}$ for different BS densities $\lambda_{\rm BS}$ under three operating protocols.}
	\vspace{-0.4cm}
\end{figure*}

Fig. 4 shows the ratio of effective serving elements for OUs and IUs as a function of the building radius $D$. As $D$ increases, the ratio decreases for both user types. Comparing Fig. 4(a) and Fig. 4(b), the ratio for OUs declines more sharply than for IUs. This is due to the fact that an OU's service depends on both the cascaded BS-CIOS and CIOS-OU links, which are more sensitive to the increasing building radius $D$.

Fig. 5 compares the successful connection probabilities of the direct, reflected, and transmitted links versus the BS density $\lambda_{\rm BS}$ for the proposed conformal deployment and the point-deployed IOS baseline in [21]. The direct-link probability is nearly identical for the two schemes, since it is independent of the surface deployment pattern. In contrast, the proposed conformal deployment achieves a markedly higher successful connection probability for both the reflected and transmitted links. For example, at ${\lambda _{{\rm{BL}}}} = 40$ ${\rm{k}}{{\rm{m}}^{ - 2}}$, the reflected link probability is about 75\% higher than that of [21], while the transmitted link probability is about 125\% higher. Moreover, the transmission performance of the baseline remains nearly unchanged for different $\lambda_{\rm BL}$, indicating its limited sensitivity to building density under point deployment. These results clearly demonstrate the advantage of conformal deployment in enlarging the feasible service region for both outdoor reflection and indoor transmission.

Fig. 6 plots the coverage probability of OUs versus the building density $\lambda_{\rm BL}$ for different BS densities $\lambda_{\rm BS}$ under three operating protocols. In Fig. 6(a), the OU coverage first increases and then decreases with $\lambda_{\rm BL}$, reflecting the tradeoff between shorter cascaded-link distances enabled by denser CIOS deployment and the intensified blockage caused by more buildings. In Fig. 6(b), the blockage effect becomes dominant due to the larger building radius, so the coverage decreases monotonically with $\lambda_{\rm BL}$. Among the three protocols, the performance difference is generally small, while MS shows a slight advantage in the moderate-blockage regime and ES becomes more favorable when $\lambda_{\rm BL}$ is large.

Fig. 7 shows the coverage probability of IUs versus the building density $\lambda_{\rm BL}$ for different BS densities $\lambda_{\rm BS}$ under three operating protocols. In both Fig. 7(a) and Fig. 7(b), the IU coverage decreases with $\lambda_{\rm BL}$, since denser buildings increase the blockage probability of the serving transmission link. The degradation becomes more pronounced for $D=20$ m, where the larger indoor region further amplifies the blockage sensitivity. Compared with ES and MS, TS is slightly less competitive in the low-blockage regime but becomes more favorable when $\lambda_{\rm BL}$ is large, indicating that transmission-oriented resource allocation is more beneficial under severe blockage.


\begin{figure*}[!t]
	\subfigure[$D=10$m]{\includegraphics[width=0.49\textwidth]{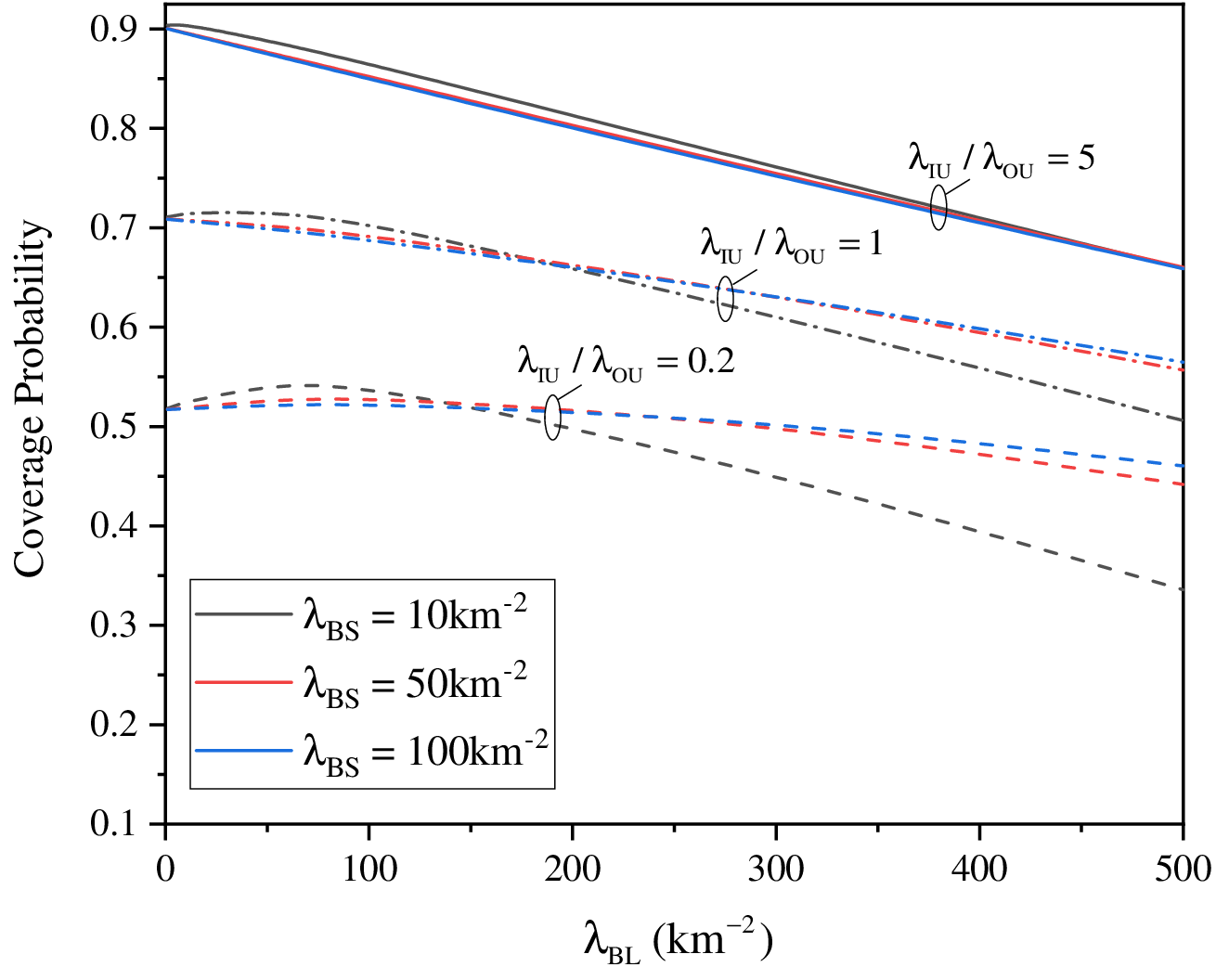}}
	\subfigure[$D=20$m]{\includegraphics[width=0.49\textwidth]{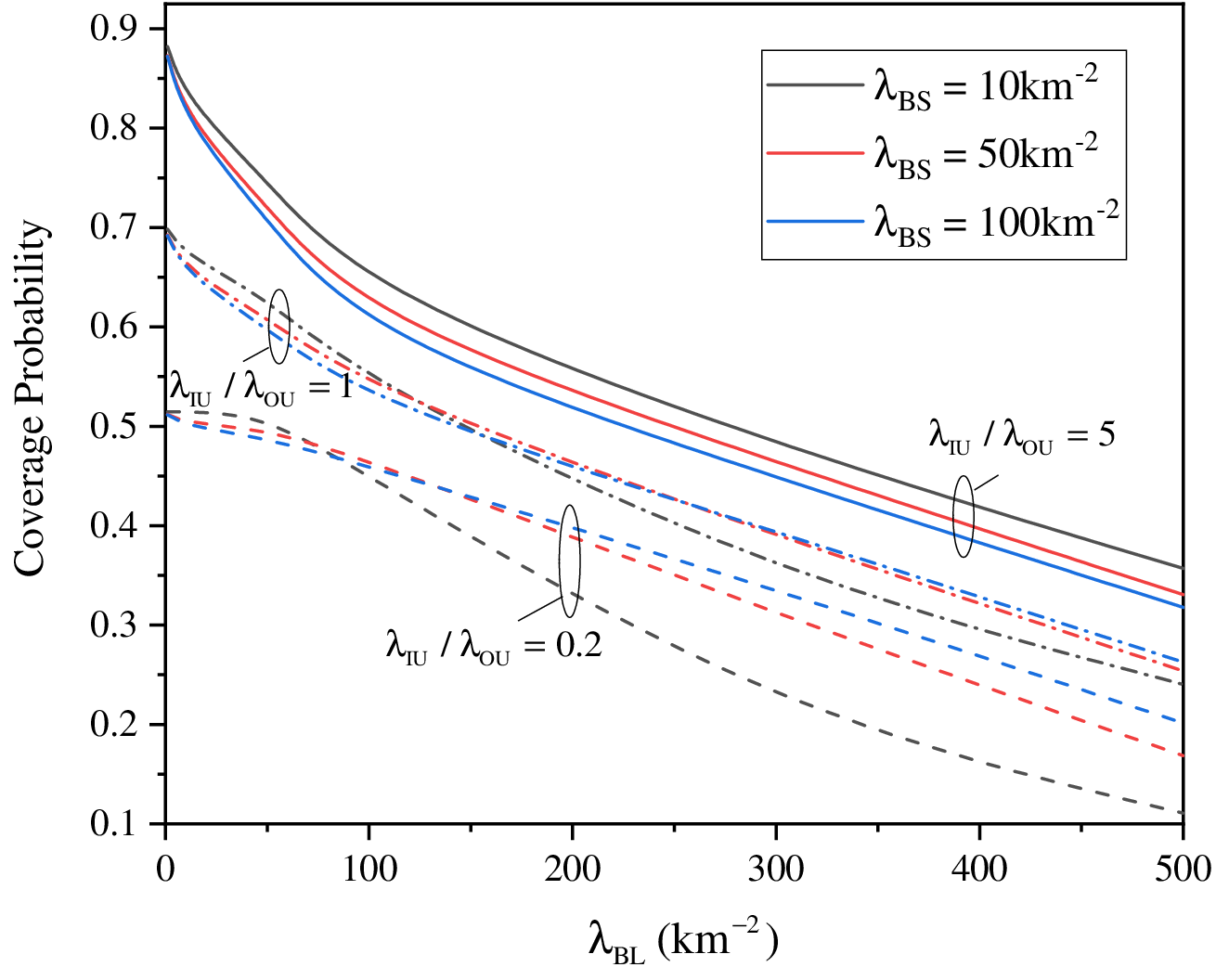}}
	\vspace{-0.3cm}
	\caption{Coverage probability versus building density $\lambda_{\rm BL}$ for different BS densities $\lambda_{\rm BS}$ and indoor/outdoor user density ratios $\lambda_{\rm IU}/\lambda_{\rm OU}$.}
		\vspace{-0.4cm}
\end{figure*}

\begin{figure*}[!t]
    \subfigure[Outdoor]{\includegraphics[width=0.33\textwidth]{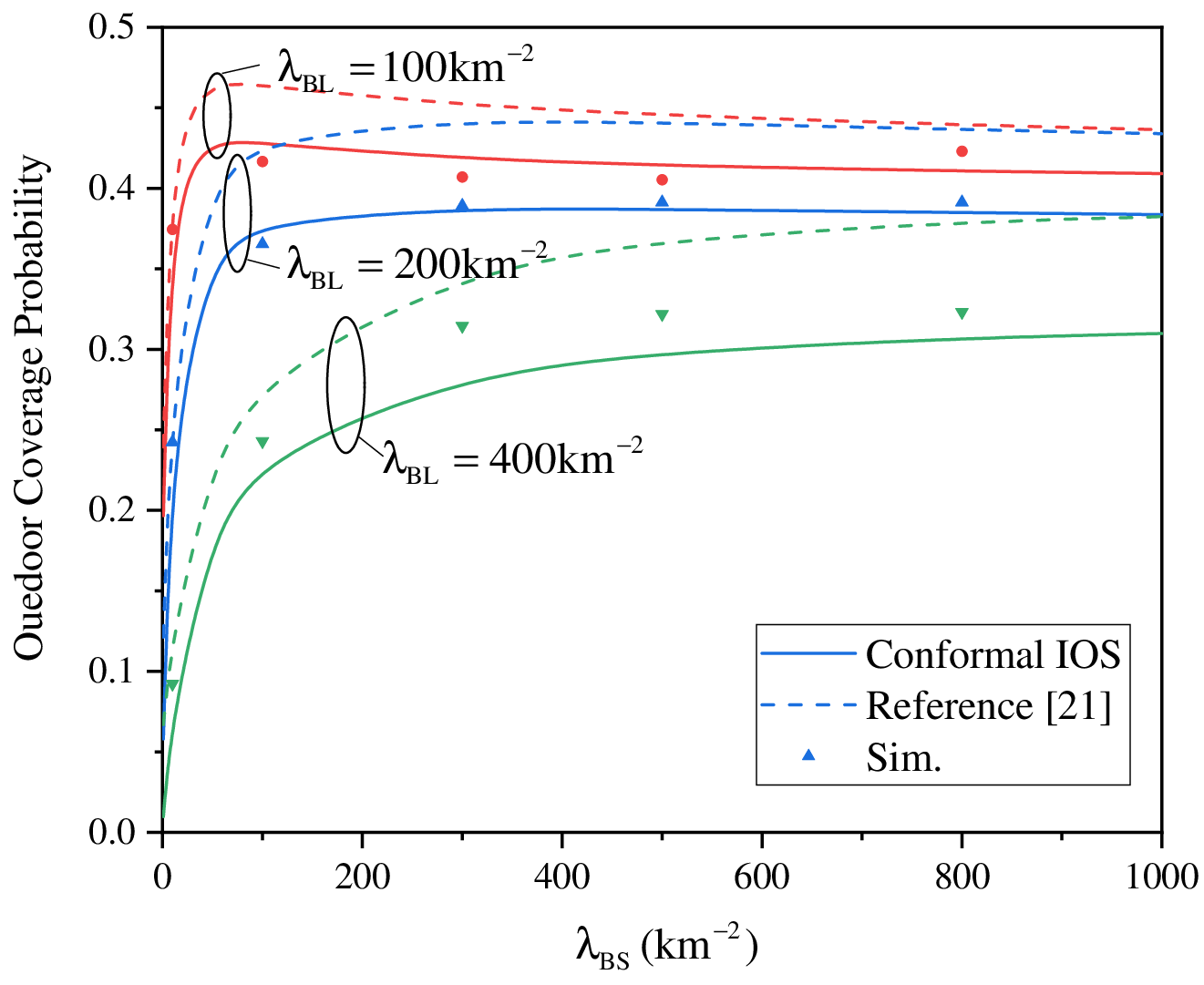}}
	\subfigure[Indoor]{\includegraphics[width=0.33\textwidth]{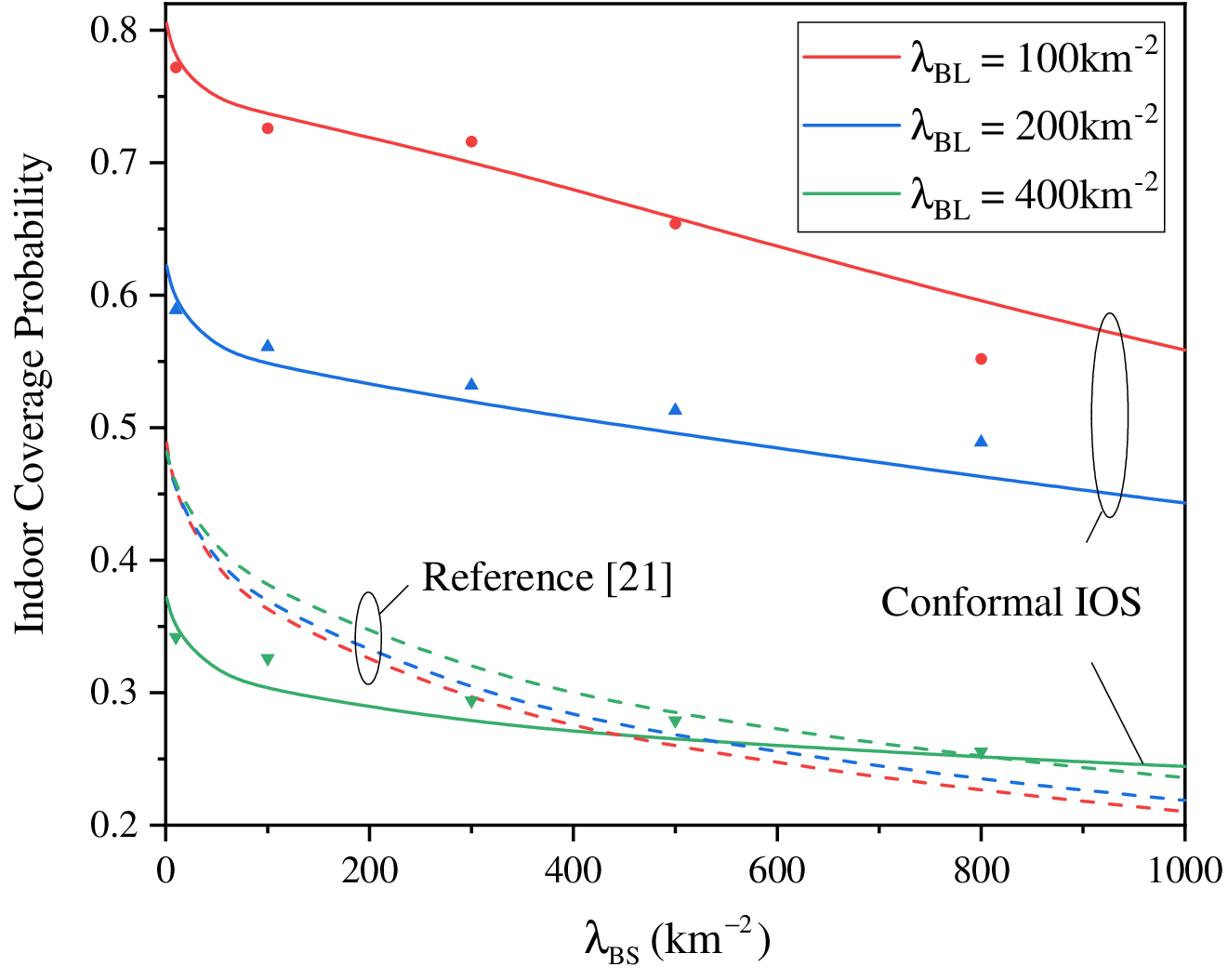}}	
	\subfigure[Overall]{\includegraphics[width=0.33\textwidth]{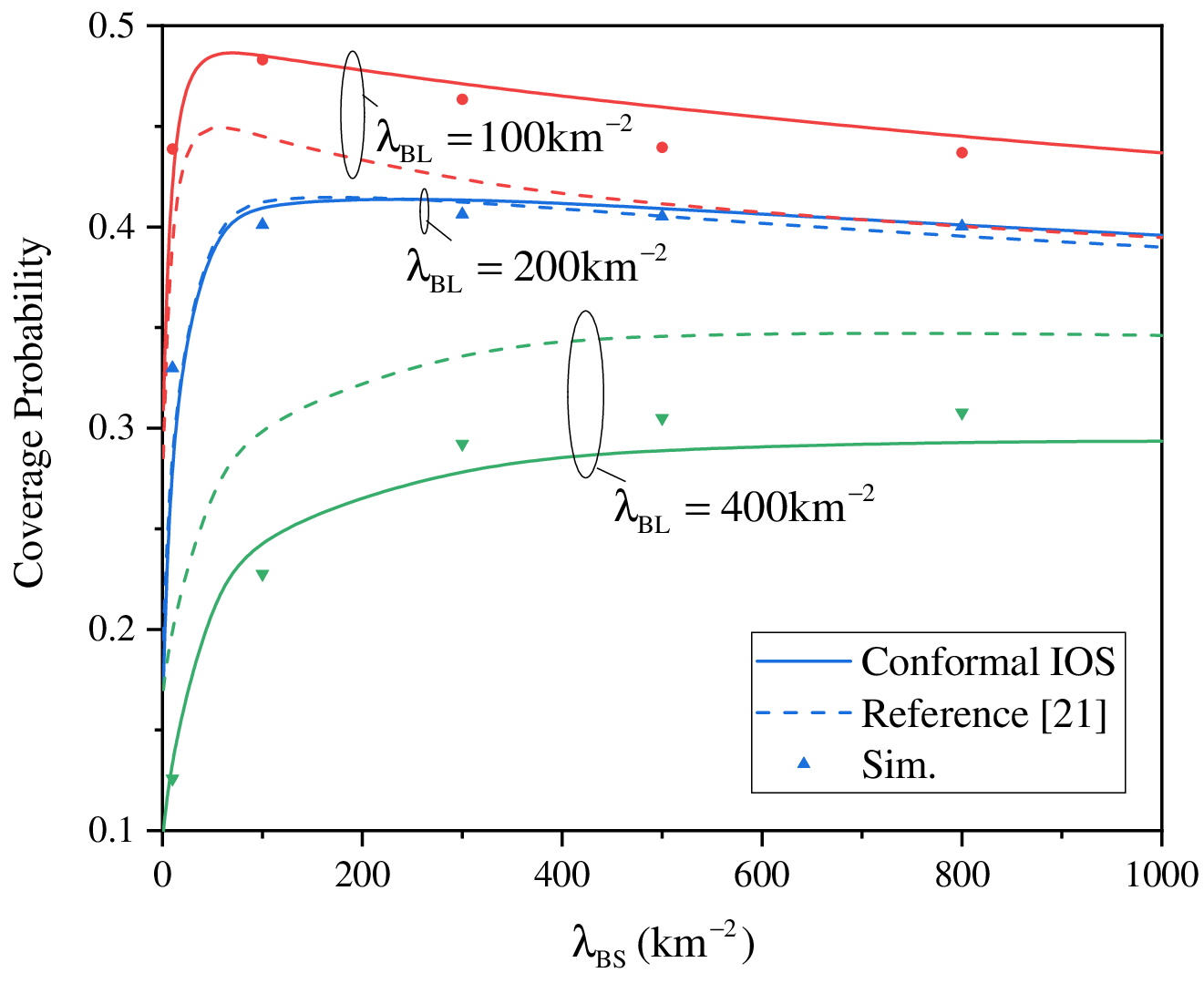}}
	\vspace{-0.5cm}
	\caption{Coverage probability versus BS density $\lambda_{\rm BS}$ for different building densities $\lambda_{\rm BL}$.}
	\vspace{-0.4cm}
\end{figure*}

\begin{figure}[!t]
	\centering	\includegraphics[width=0.49\textwidth]{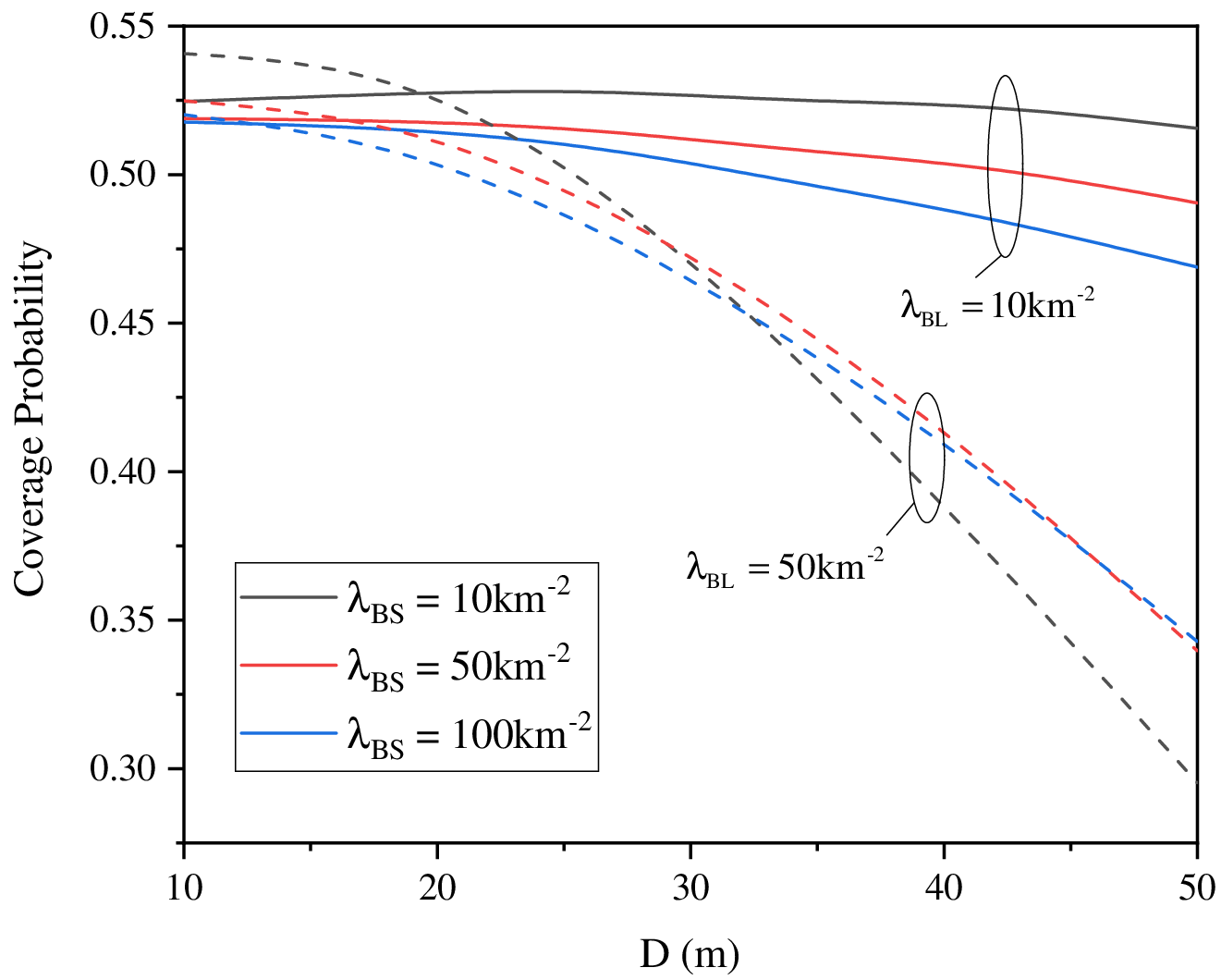}	
	\vspace{-0.4cm}
	\caption{Coverage probability versus building radius $D$ for different BS densities $\lambda_{\rm BS}$ and building densities $\lambda_{\rm BL}$.}
	\vspace{-0.4cm}
\end{figure}

Fig. 8 illustrates the overall coverage probability versus the building density $\lambda_{\rm BL}$ under different indoor to outdoor user density ratios. Since the IU coverage probability is generally higher than that of OUs, increasing the IU proportion consistently improves the overall coverage. When OUs account for a larger fraction, the overall trend becomes increase then decrease due to the balance between the distance reduction benefit brought by denser CIOS deployments and the blockage intensification caused by more buildings. As $D$ increases, the blockage dominated degradation becomes more pronounced, thereby shrinking the regime where densifying buildings yields coverage gains.

Fig. 9(a)--(c) compare the outdoor, indoor, and overall coverage probabilities versus the BS density $\lambda_{\rm BS}$ under different building densities $\lambda_{\rm BL}$ for the proposed conformal CIOS deployment and the planar IOS baseline in~[21], under the MS protocol with $\lambda_{\rm IU}/\lambda_{\rm OU}=0.2$. Compared with~[21], the proposed conformal deployment yields slightly lower outdoor coverage in Fig. 9(a). This is because, although facade curvature enlarges the feasible service region, it also reduces the effective number of participating elements, which weakens the cascaded-link gain. By contrast, Fig. 9(b) shows that the proposed scheme achieves a clear indoor-coverage advantage for sparse and moderate building densities, since conformal deployment provides a larger BS-illuminated feasible aperture for transmission. As a result, Fig. 9(c) shows that the overall coverage gain of conformal deployment is building-density dependent: conformal deployment is superior in sparse-building regimes, comparable in moderate regimes, and inferior in dense-building regimes.

Fig. 10 illustrates the overall coverage probability versus the building radius $D$ under different $\lambda_{\rm BL}$ and $\lambda_{\rm BS}$ for MS protocol. When $\lambda_{\rm BL}$ is small, the overall coverage changes only slightly with $D$, while a smaller $\lambda_{\rm BS}$ consistently yields a higher and more stable coverage due to reduced interference. 
When $\lambda_{\rm BL}$ is large, the overall coverage decreases monotonically with $D$, and the degradation is more pronounced for a smaller $\lambda_{\rm BS}$. This is because enlarging $D$ strengthens the blockage impact. Consequently, sparse BS deployments are more exposed to longer serving distances and a lower probability of successful connection, resulting in a faster coverage drop than denser BS deployments.

\vspace{-0.1cm}	
\section{Conclusion}
\vspace{-0.1cm}
This paper investigates CIOS as a facade integrated metasurface paradigm for simultaneous indoor and outdoor communication networks and developed a building coupled stochastic geometry framework for its performance analysis. A building-dependent spatial model is developed by representing buildings as a Boolean scheme of random cylinders equipped with facade-mounted CIOSs and modeling outdoor BSs as a Poisson hole process confined to the outdoor region. To capture the impact of conformal deployment, the geometry-driven service constraints are characterized and the effective number of serving elements is introduced via visible-arc metrics for both reflection and transmission. By embedding these constraints into the statistics of serving and interfering cascaded paths, the successful connection probabilities are derived and coverage probability of outdoor users, indoor users, and the overall network under different operating protocols. Numerical results show that conformal deployment significantly improves the successful connection probabilities of reflected and transmitted links, whereas its overall coverage gain depends on the geometry-induced tradeoff between service region expansion and effective aperture reduction.

\vspace{-0.2cm}	
\section*{Appendix A }
\vspace{-0.1cm}
		\section*{Proof of {Lemma 1}}
        \vspace{-0.1cm}
Conditioned on the distances $r_{{B_0}{C_0}}$ and $r_{{C_0}O}$, the visible central angles $\omega_B$ and $\omega_U$ are determined by the tangency geometry in Fig.~2(a). The corresponding visible arc intervals on the building perimeter are denoted by $\mathcal{V}_B$ and $\mathcal{V}_U$, respectively, with arc measures $|\mathcal{V}_B|=\omega_B$ and $|\mathcal{V}_U|=\omega_U$. Let

\begin{equation}
\omega_{UB}\triangleq |\mathcal{V}_U\cap\mathcal{V}_B|
\end{equation}
denote the overlap angle, which determines the effective reflection aperture.

Since the relative orientation between $\mathcal{V}_U$ and $\mathcal{V}_B$ is uniform over $[0,2\pi)$, there exist three mutually exclusive cases. First, the two visible intervals do not overlap, yielding $\omega_{UB}=0$ with probability
\begin{equation}
\Pr(\omega_{UB}=0)=1-\frac{\omega_U+\omega_B}{2\pi}.
\end{equation}

Second, one visible interval is fully contained in the other, which occurs when their starting angles fall within a window of length $|\omega_U-\omega_B|$, and thus
\begin{equation}
\Pr\!\left(\omega_{UB}=\min(\omega_U,\omega_B)\right)=\frac{|\omega_U-\omega_B|}{2\pi}.
\end{equation}

Third, the remaining probability mass corresponds to partial overlap, under which $\omega_{UB}$ is uniformly distributed over $(0,\min(\omega_U,\omega_B))$ with probability
\begin{equation}
\Pr\!\left(0<\omega_{UB}<\min(\omega_U,\omega_B)\right)=\frac{\min(\omega_U,\omega_B)}{\pi}.
\end{equation}
Therefore, the conditional distribution of $\omega_{UB}$ is a mixed distribution consisting of two atoms at $0$ and $\min(\omega_U,\omega_B)$ and a continuous uniform component on $(0,\min(\omega_U,\omega_B))$.

Under the MS protocol, only a fraction $\eta_O^{\rm MS}$ of the $N$ elements is configured for reflection
\begin{equation}
N_{\text{eff},O}=\frac{\eta_O^{\rm MS}N}{2\pi}\,\omega_{UB}.
\end{equation}
Applying this mapping to the above mixed distribution of $\omega_{UB}$ directly yields the PDF in Lemma~1.

\vspace{-0.2cm}
\section*{Appendix B }
\vspace{-0.1cm}
		\section*{Proof of {Lemma 11}}
        \vspace{-0.1cm}
Let $\zeta_i^{I}\in\{0,1\}$ indicate whether the reflected interference path from BS $i$ is active, with $\Pr(\zeta_i^{I}=1\mid \rho_i,r_i)=P_f^I(\rho_i,r_i)$. 

Due to the conformal visibility constraint, not all CIOS elements contribute to the reflected interference. Similar to the serving reflection case, only the elements lying in the intersection of the two tangency-limited visible arcs contribute. For the isosceles geometry induced by $(\rho,r)$,  the angular interval on the CIOS facade that is simultaneously visible to both the interfering BS and the OU is given by
\begin{equation}
\omega_{UB}'(\rho,r)=2\arccos\!\left(\frac{D}{d}\right)-2\arctan\!\left(\frac{r}{2\rho}\right),
\end{equation}

The reflected interference can be written as
\begin{equation}
I_r=\sum_{i\in\Phi_{\rm BS}\setminus\{0\}} \zeta_i^{I}\, g_i\, a(\rho_i,r_i),
\end{equation}
where $g_i\sim\exp(1)$. Conditioning on $r_{{B_0}O}$ and using independence across interferers,
\begin{equation}
\mathcal{L}_{I_r\mid r_{{B_0}O}}(s)
=\mathbb{E}_{\Phi_{\rm BS}}\!\left[
\prod_{i\in\Phi_{\rm BS}\setminus\{0\}}
\mathbb{E}_{g,\zeta,\rho}\!\left(e^{-s\zeta_i^{I} g\,a(\rho_i,r_i)}\right)
\Big|\,r_{{B_0}O}
\right].
\end{equation}
For a given $(\rho,r)$, $\mathbb{E}_{g,\zeta}\!\left(e^{-s\zeta g\,a(\rho,r)}\right)
=1-P_f^I(\rho,r)+\frac{P_f^I(\rho,r)}{1+s\,a(\rho,r)}$. Applying the PGFL over $\mathbb{R}^2\setminus B(\mathbf{U},r_{{B_0}O})$ yields \eqref{eq:laplace_reflected_interference_opt}.

\vspace{-0.2cm}
\section*{Appendix C}
\vspace{-0.1cm}
		\section*{Proof of {Lemma 14}}
        \vspace{-0.1cm}
Let $r_i$ denote the distance from interfering BS $i$ to the host building center $L_0$, and let $\varphi_i$ be the angle between the directions of BS $i$ and the IU with respect to $L_0$. Then the corresponding IU-side distance is $\sqrt{r_i^2+r_{{L_0}I}^2-2r_i r_{{L_0}I}\cos\varphi_i}$, which yields $\ell_I(r_i,\varphi_i)$.

Not all facade elements can be simultaneously illuminated by BS $i$ and contribute to the IU reception. By the same tangency geometry as for the serving transmission link, the effective aperture for an interferer at distance $r$ is captured by $N_{\text{eff},I}'(r)$ in \eqref{eq:NeffI'}. Let $\zeta_i^{t}\in\{0,1\}$ indicate whether the corresponding transmission interference path is unblocked, with $\Pr(\zeta_i^{t}=1\mid r_i)=P_s^I(r_i)$.

The aggregate transmission interference is
\begin{equation}
I_t=\sum_{i\in\Phi_{\rm BS}\setminus\{0\}} \zeta_i^{t}\, \eta_I^{\rm ES}\,N_{\text{eff},I}'(r_i)\, g_i\,\ell_I(r_i,\varphi_i),
\end{equation}
where $g_i\sim\exp(1)$. Conditioning on $(r_{{B_0}{L_0}},r_{{L_0}I})$ and using independence,
\begin{equation}
\mathbb{E}_{g,\zeta}\!\!\left[\!e^{-s\!\zeta\!\,\!\eta_I^{\rm ES} \!N_{\text{eff},I}'\!(\!r\!)\!\,\!g\!\,\ell_I\!(r,\varphi)}\!\!\right]\!
\!=\!\!1\!-\!P_s^I(r)+\frac{P_s^I(r)}{1\!+\!s\!\,\eta_I^{\rm ES}\! N_{\text{eff},I}'\!(r)\!\,\ell_I\!(r,\varphi)}.
\end{equation}

Since BS$_0$ is the nearest outdoor BS to $L_0$, all interferers satisfy $r\ge r_{{B_0}{L_0}}$. Applying the PGFL over $\mathbb{R}^2\setminus B(L_0,r_{{B_0}{L_0}})$, together with the uniformity of $\varphi$, yields \eqref{eq:laplace_transmission_interference}.

\end{document}